\documentclass[usenatbib,hyperpdf]{mn2e}
\usepackage{graphicx}

\def\ni{\noindent}                                    %No indent%
\def\etal{et\thinspace al.\thinspace}                    %et al.%
\def\starlight{\textsc{starlight}}                    %Starlight%
\title[The history of star-forming galaxies in the SDSS] 
{The history of star-forming galaxies in the Sloan Digital Sky
Survey}

\author[Asari \etal]
        {N. V. Asari$^{1}$,
	R. Cid Fernandes$^{1}$,
	G. Stasi\'nska $^{2}$,
	J. P. Torres-Papaqui$^{1,3}$,
  	A. Mateus$^{4}$,
	\newauthor
	L. Sodr\'e Jr.$^{5}$,
	W. Schoenell$^{1}$,
	J. M. Gomes$^{1}$
	(the SEAGal collaboration)\thanks{Semi-Empirical Analysis of Galaxies}\\
	$^{1}$Departamento de F\'{\i}sica - CFM - Universidade Federal de
        Santa Catarina, Florian\'opolis, SC, Brazil\\
	$^{2}$LUTH, Observatoire de Paris, CNRS, Universit\'e Paris
	Diderot; Place Jules Janssen 92190 Meudon, France\\
	$^{3}$Instituto Nacional de Astrof\'{\i}sica, \'Optica y
	Electr\'onica, Apdo. Postal 51 y 216, 72000 Puebla, M\'exico\\
	$^{4}$Instituto de Ciencias del Espacio (IEEC-CSIC),
        Barcelona, Spain\\
	$^{5}$Instituto de Astronomia, Geof\'{\i}sica e Ci\^encias
        Atmosf\'ericas, Universidade de S\~ao Paulo, S\~ao Paulo, SP,
        Brazil\\
        }

\begin{document}

\maketitle

%***************************************************************%
%                                                               %
%                           Abstract                            %
%                                                               %
%***************************************************************%

\begin{abstract} 
This paper, the 6$^{\rm th}$ in the Semi-Empirical Analysis of
Galaxies series, studies the evolution of 82302 star-forming (SF)
galaxies from the Sloan Digital Sky Survey. Star formation histories
(SFH) are derived from detailed spectral fits obtained with our
publicly available spectral synthesis code \starlight.  Our main goals
are to explore new ways to derive SFHs from the synthesis results and
apply them to investigate how SFHs vary as a function of nebular
metallicity ($Z_{neb}$).  A number of refinements over our previous
work are introduced, including (1) an improved selection criterion;
(2) a careful examination of systematic residuals around H$\beta$; (3)
self-consistent determinations of nebular extinctions and
metallicities; (4) tests with several $Z_{neb}$ estimators; (5) a
study of the effects of the reddening law adopted and of the relation
between nebular and stellar extinctions and the interstellar component
of the NaI D doublet.

Our main achievements may be summarized as follows: (1) A conventional
correlation analysis is performed to study how global properties
relate to $Z_{neb}$, leading to the confirmation of previously known
relations, such as those between $Z_{neb}$ and galaxy luminosity,
mass, dust content, mean stellar metallicity and mean stellar age.
(2) A simple formalism which compresses the results of the synthesis
while at the same time yielding time dependent star formation rates
(SFR) and mass assembly histories is presented. (3) A comparison of
the current SFR derived from the population synthesis with that
obtained from H$\alpha$ shows that these independent estimators agree
very well, with a scatter of a factor of two. An important corollary
of this finding is that we now have a way to estimate SFR in galaxies
hosting AGN, where the H$\alpha$ method cannot be applied.  (4) Fully
time dependent SFHs were derived for all galaxies, and then averaged
over six $Z_{neb}$ bins spanning the entire SF wing in the ${\rm
[OIII]/H}\beta \times {\rm [NII]/H}\alpha$ diagram. (5) We find that
SFHs vary systematically along the SF sequence.  Though all
star-forming galaxies formed the bulk of their stellar mass over 1 Gyr
ago, low $Z_{neb}$ systems evolve at a slower pace and are currently
forming stars at a much higher relative rate.  Galaxies at the tip of
the SF wing have current specific SFRs about 2 orders of magnitude
larger than the metal rich galaxies at its bottom. (6) At any given
time, the distribution of specific SFRs for galaxies within a
$Z_{neb}$-bin is broad and approximately log-normal. (7) The whole
study was repeated grouping galaxies within bins of stellar mass and
surface mass density, both of which are more fundamental drivers of
SFH.  Given the existence of strong
$Z_{neb}$--$M_\star$--$\Sigma_\star$ relations, the overall picture
described above remains valid. Thus, low $M_\star$ (low
$\Sigma_\star$) systems are the ones which evolves slower, with
current specific SFRs much larger than more massive (dense) galaxies.
(8) This overall pattern of SFHs as a function of $Z_{neb}$, $M_\star$
or $\Sigma_\star$ is robust against changes in selection criteria,
choice of evolutionary synthesis models for the spectral fits, and
differential extinction effects.
\end{abstract}


%***************************************************************%
%                                                               %
%                        Introduction                           %
%                                                               %
%***************************************************************%

\section{Introduction}
\label{sec:Introduction}

The Sloan Digital Sky Survey (SDSS, \citealp{York_etal_2000}), with
its homogeneous spectroscopic and photometric data on hundreds of
thousands of galaxies has revolutionized our perception of the world
of galaxies in the local Universe. The enormous amount of objects
allowed one to reveal trends that had not been suspected before. For
example, while is was known since the work of
\citet*{Baldwin_Phillips_Terlevich_1981} that objects ionized by
massive stars and active galactic nuclei (AGN) live in different zones
of emission line ratios diagrams, the fact that emission line galaxies
are distributed in two well defined wings
(\citealp{Kauffmann_etal_2003c}) in the famous [OIII]5007/H$\beta$ vs
[NII]6583/H$\alpha$ diagnostic diagram (heareafter, the BPT diagram)
came as a surprise.

The left wing of the BPT diagram can be understood as a sequence in
metallicity of normal star-forming (SF) galaxies. The present-day
nebular metallicity ($Z_{neb}$) of a galaxy is intimately connected
with its past star formation history (SFH). The main goal of this
paper is to explore this link. The focus of many of the pioneering
studies of SFH of SF galaxies was instead the variation of SFH with
the Hubble type. \citet*{Searle_Sargent_Bagnuolo_1973}, for instance,
assumed a simple model for the SFH and calculated $UBV$ colors for
simulated galaxies. Comparing simulated and observed colors, they
concluded that morphological type alone does not explain the
differences in SFH, proposing that the galaxy mass should also be used
as a tracer of star formation.

\citet*{Gallagher_Hunter_Tutukov_1984} introduced a way to study the
star formation rates (SFR) in three different epochs of a galaxy's
history.  In order to achieve such time resolution, manifold indices
were used: HI observations, dynamical masses, $B$-band and H$\alpha$
luminosities, and $UBV$ colors.  \citet{Sandage_1986} applied some of
the techniques presented by \citet{Gallagher_Hunter_Tutukov_1984} to
investigate differences in SFH along the the Hubble sequence.  In the
same vein, \citet*{Kennicutt_Tamblyn_Congdon_1994} derived the SFR for
SF objects from the H$\alpha$ luminosity and $UBV$ colors, and found
that the SFH differences for galaxies of the same Hubble type have a
much stronger relation with their disk than with their bulge.
\citet{Gavazzi_etal_2002} measured the present and past SFRs of
late-type galaxies in nearby clusters from H$\alpha$ imaging and near
infrared observations, and also derived the global gas content from HI
and CO observations.  Most of these studies had to rely on many
different indices and observations in order to measure an
instantaneous SFR, or at most a 2--3 age resolution star formation
history for SF galaxies.

\citet{Bica_1988} introduced a method to reconstruct SFHs in greater
detail by mixing the properties of a base of star clusters of various
ages ($t_\star$) and metallicities ($Z_\star$).  In its original
implementation, this method used a set of 5--8 absorption line
equivalent widths as observables, a grid of clusters arranged in 35
combinations of $t_\star$ and $Z_\star$, and a simple parameter space
exploration technique limited to paths through the $t_\star$-$Z_\star$
plane constrained by chemical evolution arguments (see also
\citealp{Schmidt_etal_1991, Bica_Alloin_Schmitt_1994,
CidFernandes_etal_2001}). Its application to nuclear spectra of nearby
galaxies of different types revealed systematic variations of the SFH
along the Hubble sequence.

For over a decade the most attractive feature of Bica's method was
its use of observed cluster properties, empirically bypassing the
limitations of evolutionary synthesis models, which until recently
predicted the evolution of stellar systems at spectral resolutions
much lower than the data.  This is no longer a problem. Medium and
high spectral resolution stellar libraries, as well as updates in
evolutionary tracks have been incorporated into evolutionary synthesis
models in the past few years. The current status of these models and
their ingredients is amply discussed in the proceedings of the IAU
Symposium 241 (\citealp{Vazdekis_Peletier_2007}).

These advances spurred the development of SFH recovery methods which
combine the non-parametric mixture approach of empirical population
synthesis with the ambitious goal of fitting galaxy spectra on a
pixel-by-pixel basis using a base constructed with these new
generation of evolutionary synthesis models.  Methods based on
spectral indices have also benefitted from these new models, and
produced an impressive collection of results (e.g.,
\citealp{Kauffmann_etal_2003a, Kauffmann_etal_2003b,
Kauffmann_etal_2003c, Gallazzi_etal_2005, Brinchmann_etal_2004}).
However, current implementations of these methods do not reconstruct
detailed SFHs, although they do constrain it, providing estimates of
properties such as mass-to-light ratios, mean stellar age, fraction of
mass formed in recent bursts, and ratio of present to past SFR.

The first SFHs derived from full spectral fits of SDSS galaxies were
carried out with the MOPED (\citealp{Panter_etal_2003_Moped,
Panter_etal_2007_Moped}; \citealp*{Mathis_Charlot_Brinchmann_2006})
and \starlight\ codes. MOPED results have been recently reviewed by
Panter (2007), so we just give a summary of the results achieved with
\starlight.

\starlight\ itself was the main topic of the first paper in our
Semi-Empirical Analysis of Galaxies series (SEAGal).  In SEAgal I
(\citealp{CidFernandes_etal_2005_SEAGal1}) we have thoroughly
evaluated the method by means of simulations, astrophysical
consistency tests and comparisons with the results obtained by
independent groups.  In SEAGal II (\citealp{Mateus_etal_2006_SEAGal2})
we have revisited the bimodality of the galaxy population in terms of
spectral synthesis products.  In SEAGal III
(\citealp{Stasinska_etal_2006_SEAGal3}), we combined the emission
lines dug out and measured from the residual spectrum obtained after
subtraction of the synthetic spectrum with photoionization models to
refine the criteria to distinguish between normal SF galaxies and AGN
hosts. SEAGal IV (\citealp{Mateus_etal_2007_SEAGal4}) deals with
environment effects, studied in terms of the relations between mean
age, current SFR, density, luminosity and mass.

Only in SEAGal V (\citealp{CidFernandes_etal_2007_SEAGal5}) we turned
our attention to the detailed time dependent information provided by
the synthesis.  We have used the entire Data Release 5
(\citealp{Adelman-McCarthy_etal_2007}) to extract the population of SF
galaxies, and study their chemical enrichment and mass assembly
histories. It was shown that there is a continuity in the evolution
properties of galaxies according to their present properties: Massive
galaxies formed most of their stars very early and quickly reached the
high stellar metallicities they have today, whereas low mass (metal
poor) galaxies evolve slower. These findings are in agreement with
recent studies of the mass assembly of larges samples of galaxies
through the fossil record of their stellar populations
(\citealp{Heavens_etal_2004_Moped}), and of studies of the $Z_\star$
distribution in small samples of galaxies (e.g.,
\citealp{Skillman_Cote_Miller_2003} and references therein), but the
generality of the result applied to the entire population of SF
galaxies was shown for the first time.

In the present paper, we aim at a more complete view of the properties
of SF galaxies, and their variations along the SF sequence in the BPT
diagram, improving and expanding upon the results only briefly
sketched in SEAGal V.  In particular, we discuss in depth time
averaged values of quantities such as the SFR and the SFR per unit
mass, as well as their \emph{explicit} time dependence.

The paper is organized as follows.  Section 2 describes our parent
sample and explain our criteria to define normal SF galaxies. This
section also explains how we deal with extinction and how we estimate
the nebular metallicity. In Section 3, we discuss the global
properties of galaxies along the SF sequence in the BPT diagram. In
Section 4, we then proceed to explain our formalism to uncover the
explicit time dependence of such quantities as the star formation
rate. In Section 5, we show that the current star formation rate as
estimated by the most commonly indicator -- the H$\alpha$
luminosity -- compares with that obtained from our stellar population
synthesis analysis. In Section 6, we analyse the SFH along the SF
sequence, binning galaxies in terms of their present-day nebular
metallicity. We show that, despite the important scatter at any
$Z_{neb}$, there is a clear tendency for the SFH as a function of
$Z_{neb}$, in that in the most metal-rich galaxies most of the stellar
mass assembly occurred very fast and early on, while metal poor
systems are currently forming stars at much higher relative rates.  
%Although, qualitatively, this result has been known for long
%(e.g., \citealp{Lequeux_etal_1979, Sandage_1986, Bica_1988,
%Skillman_etal_1989, Zaritsky_etal_1994, Melbourne_Salzer_2002,
%Lamareille_etal_2004}), {\bf ojo Change REFS! Estas sao na maioria
%sobre a M-Z relation!} it is the first time that a quantitative and
%fully time dependent analysis is presented.
We also compute mean SFHs binning the galaxies with respect to the
stellar mass and the surface mass density, which are expected to
better express the causes of the evolution of galaxies. Section
\ref{sec:samples} discusses possible selection effects and other
caveats.  Finally, Section \ref{sec:Summary} summarizes our main
results.

%***************************************************************%
%                                                               %
%                            Data                               %
%                                                               %
%***************************************************************%

\section{Data}
\label{sec:Data}
\label{sec:BPTdiagram}

%---------------------------- Figure ----------------------------
\begin{figure*}
   \includegraphics[width=0.815\textwidth, bb=60 170 592 718]{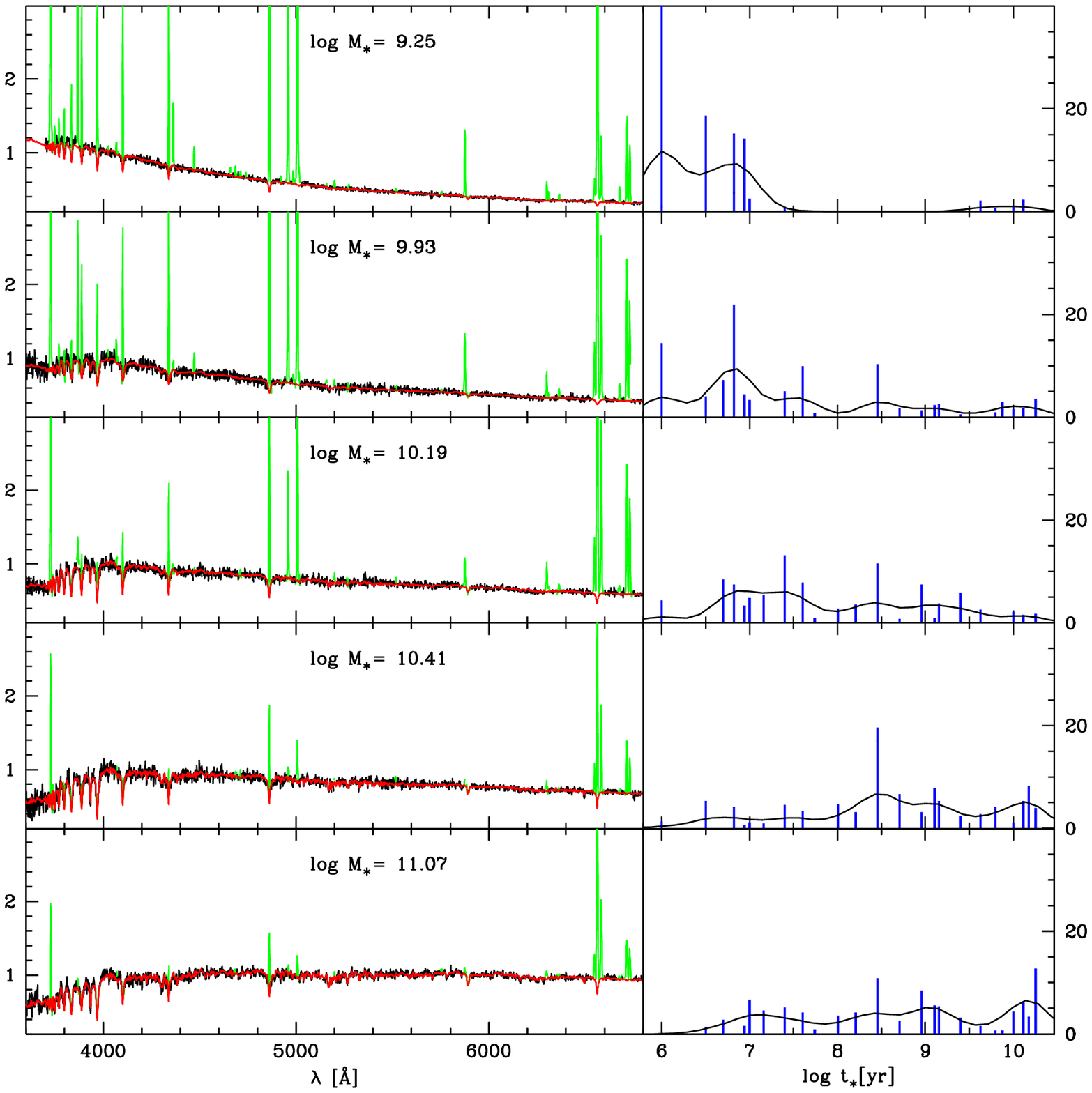}
   \includegraphics[width=0.150\textwidth, bb=0 -37 113 563]{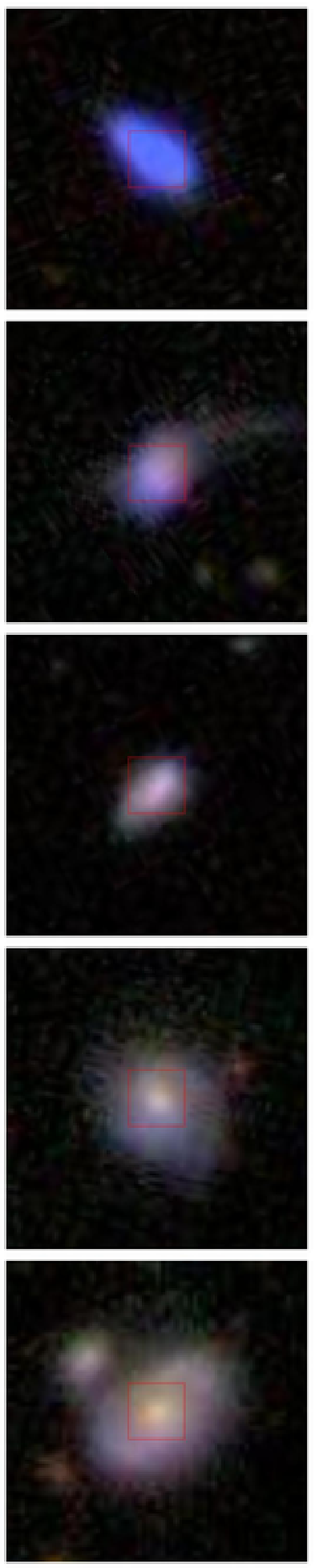}
   \caption{Five examples of the spectral fits. Left panels show the
     observed (black) and fitted (red) spectra, both normalized at
     $\lambda_0 = 4020$ \AA. Magenta lines mark regions not used in
     the fits either because they contain emission lines or because of
     artifacts in the data. Middle panels illustrate the fraction of
     light at $\lambda_0$ associated to each of the 25 SSP ages used
     in the fits.  Curves represent a 0.5 dex smoothed version of the
     population vector.  Right panels show SDSS $25.6^{\prime\prime}
     \times 25.6^{\prime\prime}$ images ($\sim$ $12 \times 12$--$34
     \times 34$ kpc$^2$).  Galaxies in this plot are ordered according
     to their nebular metallicity ($Z_{neb}$; see Section
     \ref{sec:Z_neb}). From top to bottom, $Z_{neb} = 0.29$, 0.43,
     0.61, 0.84 and 0.97 $Z_\odot$.}
\label{fig:STARLIGHT_fits}
\end{figure*}
%---------------------------- Figure ----------------------------

The data analysed in this work was extracted from the SDSS Data
Release 5 (DR5; \citealp{Adelman-McCarthy_etal_2007}). This release
contains data for 582471 objects spectroscopically classified as
galaxies, from which we have found $\sim 1.6$ per cent of duplicates,
that is, objects with multiple spectroscopic information in the parent
galaxy catalog.

From the remaining 573141 objects we have selected our parent sample
adopting the following selection criteria: $14.5 \le m_r \le 17.77$
and $z \ge 0.002$. The magnitude range comes from the definition of
the Main Galaxy Sample, whereas the lower redshift limit is used to
avoid inclusion of intragalactic sources.  The resulting sample
contains 476931 galaxies, which corresponds to about 82 per cent of
all galaxies with spectroscopic data gathered by SDSS and publicly
available in the DR5. These limits imply a reduction by $\sim 17\%$ in
the sample studied in SEAGal V.

\subsection{STARLIGHT fits}
\label{sec:STARLIGHTfits}

After correcting for Galactic extinction (with the maps of
\citealp*{Schlegel_Finkbeiner_Davis_1998} and the reddening law of
\citealp*{Cardelli_Clayton_Mathis_1989}, using $R_V = 3.1$), the
spectra were shifted to the rest-frame, resampled to $\Delta \lambda =
1$ \AA\ between 3400 and 8900 \AA, and processed through the
\starlight\ spectral synthesis code described in SEAGal I and II.

\starlight\ decomposes an observed spectrum in terms of a sum of
simple stellar populations (SSPs), each of which contributes a
fraction $x_j$ to the flux at a chosen normalization wavelength
($\lambda_0 = 4020$ \AA). As in SEAGal II, we use a base of $N_\star =
150$ SSPs extracted from the models of
\citet[BC03]{Bruzual_Charlot_2003_BC03}, computed for a
\citet{Chabrier_2003} initial mass function (IMF), ``Padova 1994''
evolutionary tracks (\citealp{Alongi_etal_1993, Bressan_etal_1993,
Fagotto_etal_1994a, Fagotto_etal_1994b, Girardi_etal_1996}), and
STELIB library (\citealp{LeBorgne_etal_2003}). The base components
comprise 25 ages between $t_{\star,j} = 1$ Myr and 18 Gyr, and 6
metallicities, from $Z_{\star,j} = 0.005$ to 2.5 solar. Bad pixels,
emission lines and the NaD doublet are masked and left out of the
fits. The emission line masks were constructed in a galaxy-by-galaxy
basis, following the methodology outlined in SEAGal II and
\citet{Asari_2006}. \starlight\ outputs several physical properties,
such as the present-day stellar mass, stellar extinction, mean stellar
ages, mean metallicities as well as full time dependent star formation
and chemical evolution histories, which will be used in our analysis.
Section \ref{sec:STARLIGHT} describes aspects of the code relevant to
this work.

Fitting half a million spectra represented a massive computational
effort, carried out in a network of over 100 computers spread over 3
continents and controlled by a specially designed PHP code. This huge
d'atabase of spectral fits and related products, as well as \starlight\
itself, are {\em publicly available} in a Virtual Observatory
environment at www.starlight.ufsc.br (see \citealp[in
prep.]{CidFernandes_etal_2007b}).

Examples of the spectral fits obtained for 5 star-forming galaxies are
shown in Fig \ref{fig:STARLIGHT_fits}.  We have ordered the galaxies
according to their nebular metallicity ($Z_{neb}$), as defined in
Section \ref{sec:Z_neb}, to illustrate how spectral characteristics
change along the $Z_{neb}$ sequence.  Metal-poor galaxies (top) show
blue spectra and strong emission lines in comparison to the redder
spectra and weaker emission lines of galaxies with a metal-rich ISM
(bottom).  The middle panels in Fig \ref{fig:STARLIGHT_fits} show the
fractional contribution to the total flux at $\lambda_0 = 4020$ \AA\
of simple stellar population (SSP) of age $t_\star$.  These panels
show that young stellar populations make a dominant contribution in
galaxies with low $Z_{neb}$, whereas at higher nebular metallicities a
richer blend of stellar ages is present.

\subsection{Emission lines}

\subsubsection{General procedure}

Emission lines were measured fitting gaussians to the residual spectra
obtained after subtraction of the stellar light using an updated
version of the line-fitting code described in SEAGal III. The main
transitions used in this study are H$\beta$, [OIII]$\lambda5007$,
H$\alpha$ and [NII]$\lambda6584$. In the next section these lines are
used to define the sub-sample of star-forming galaxies which will be
studied in this paper.

\subsubsection{The special case of H$\beta$}

We find that a zero level residual continuum is adequate to fit the
emission lines, except for H$\beta$. Inspection of the spectral fits
shows that the synthetic spectrum is often overestimated in the
continuum around H$\beta$, creating a broad, $\sim 200$ \AA\ wide
absorption trough in the residual spectrum. This problem, which can
hardly be noticed in Fig \ref{fig:STARLIGHT_fits}, becomes evident
when averaging many residual spectra (see SEAGal V and
\citealp{Panter_etal_2007_Moped}), and tends to be more pronounced for
older objects. The comparison between STELIB stars and theoretical
models presented by \citet{Martins_etal_2005} gives a clue to the
origin of this problem. A close inspection of Fig 21 in their paper
shows that the STELIB spectrum has an excess of flux on both sides of
H$\beta$ when compared to the model spectrum. This suggests that the
``H$\beta$ trough'' is related to calibration issues in the STELIB
library in this spectral range. This was confirmed by \starlight\
experiments which showed that the problem disappears using the SSP
spectra of \citet[based on the Martins \etal 2005
library]{Gonzalez-Delgado_etal_2005} or those constructed with the
MILES library (\citealp{Sanchez-Blazquez_etal_2006}).\footnote{We
thank Drs.\ Enrique Perez, Miguel Cervi\~no and Rosa
Gonz\'alez-Delgado for valuable help on this issue.}

Though this is a low amplitude mismatch (equivalent width $\sim 3$
\AA\ spread over $\sim 200$ \AA), it makes H$\beta$ sit in a region of
negative residual flux, so assuming a zero level continuum when fitting
a gaussian may chop the base of the emission line, leading to an
underestimation of its flux.  To evaluate the magnitude of this effect
we have repeated the H$\beta$ fits, this time adjusting the continuum
level from two side bands (4770--4830 and 4890--4910 \AA). On average,
the new flux measurement are 2\% larger than the ones with the
continuum fixed at zero. The difference increases to typically 4\% for
objects with $W_{H\beta} < 5$ \AA, but $S/N > 3$ in the line, and 7\%
for $W_{H\beta} < 2$ \AA.  Noise in the side bands introduces
uncertainties in the measurement of the flux, but at least it removes
the systematic effect described above, so the new measurements should
be considered as more accurate on average.  We adopt these new
H$\beta$ measurements throughout this work. Using the zero continuum
measurements changes some of the quantitative results reported in this
paper minimally, with no impact on our general conclusions.

%***************************************************************%
%                                                               %
%                        The BPT diagram                        %
%                                                               %
%***************************************************************%

\subsection{Definition of the Star Forming Sample}
\label{sec:sample_definition}

%---------------------------- Figure ----------------------------
\begin{figure}
   \includegraphics[width=0.5\textwidth]{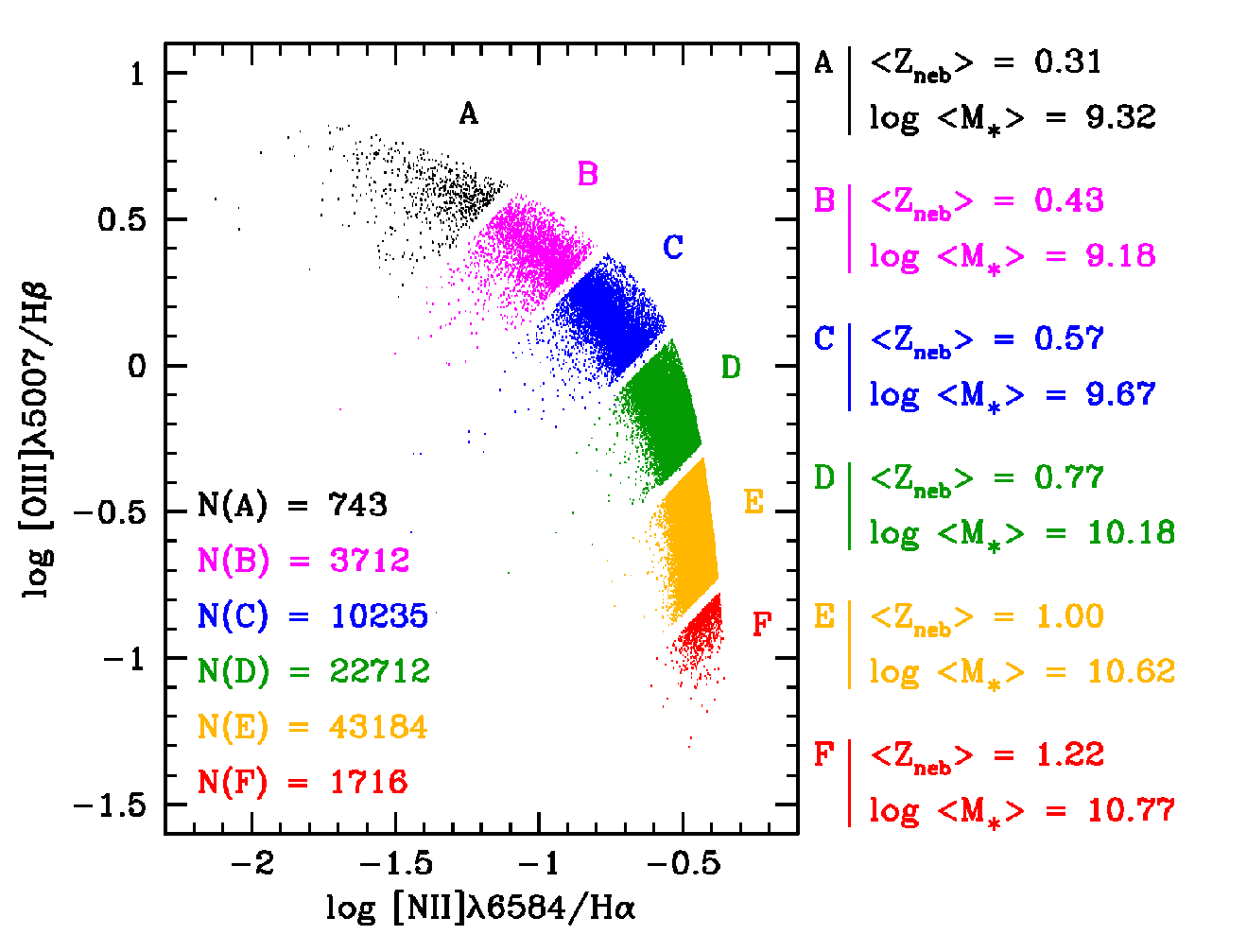}
   \caption{The SF sample in the BPT plane, chopped into bins of
     nebular abundance $Z_{\rm neb} = \frac{({\rm O/H})}{({\rm
     O/H})_\odot}$. All lines have been corrected by reddening (see
     Section \ref{sec:AV_neb}).  These same bins are used throughout
     this paper. The number of galaxies in each bin is given on the
     left. On the right, we show the corresponding mean $Z_{neb}$ and
     log mean $M_\star$ values (in solar units). Galaxies close to bin
     borders are not plotted for clarity. }
\label{fig:BPT}
\end{figure}
%---------------------------- Figure ----------------------------

Since the pioneering work of \citet*{Baldwin_Phillips_Terlevich_1981},
emission line objects are classified in terms of their location in
diagrams involving pairs of line ratios. As explained in SEAGal III,
the [NII]$\lambda6584$/H$\alpha$ vs.\ [OIII]$\lambda5007$/H$\beta$
diagram (the BPT diagram) is the most useful for this purpose, mainly
due to the partially secondary nature of N (i.e., the increase of N/O
as O/H increases, e.g., \citealp{Liang_etal_2006, Molla_etal_2006}).

Our sample of star-forming galaxies is composed by objects which are
below the line separating normal star-forming galaxies and AGN hosts
proposed in SEAGal III. We have imposed a lower limit of 3 in $S/N$ on
the 4 lines in the BPT diagram, and $S/N \ge 10$ in the 4730--4780
\AA\ continuum to constitute our main sample (the SF sample).  The
82302 galaxies composing the SF sample are shown in Fig.
\ref{fig:BPT} on the BPT plane.

Both the \starlight\ fits (and thus all SFH-related parameters) and the
emission line data are affected by the quality of the spectra. To
monitor this effect, we have defined a ``high-quality'' sub-set (the
SF$^{hq}$ sample) by doubling the $S/N$ requirements for the
SF sample, i.e., $S/N \ge 6$ in all 4 lines in the BPT diagram and a
continuum $S/N$ of 20 or better.  A total of 17142 sources satisfy
these criteria.

\begin{figure*}
   \includegraphics[bb= 70 570 572 690,width=\textwidth]
		  {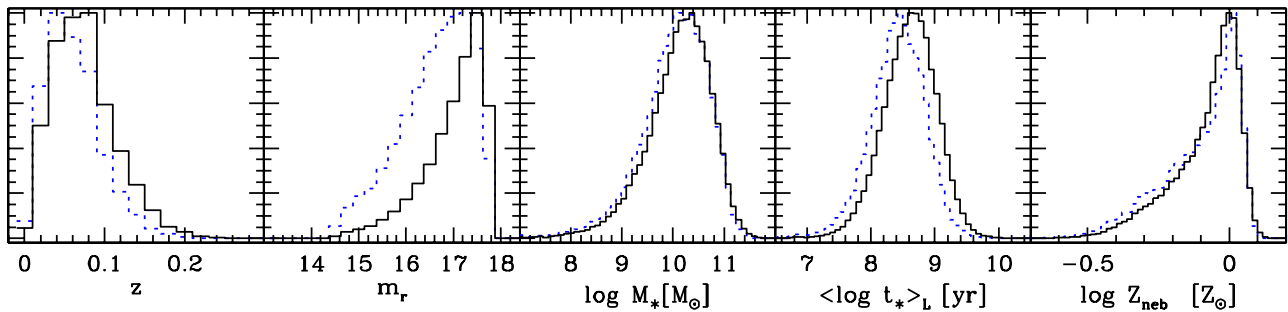}
   \caption{Normalized histograms of observed and physical properties
     for the SF (solid), SF$^{hq}$ (dotted) samples.}
\label{fig:obs_prop}
\end{figure*}

Fig. \ref{fig:obs_prop} shows the distributions of observational and
physical properties for the samples. Naturally, the SF$^{hq}$ sample
is skewed towards closer and brighter galaxies with respect to the
SF sample, but in terms of physical properties such as stellar mass,
mean age and nebular metallicity the two samples are similar.

%---------------------------- Figure ----------------------------
%---------------------------- Figure ----------------------------

\subsection{Nebular Metallicity Estimate}
\label{sec:Z_neb}

It is well known that the SF-wing in the BPT diagram is a sequence in
nebular metallicity (SEAGal III and references therein), which we
quantify by the oxygen abundance obtained through the O$_3$N$_2 =$
[OIII]5007/[NII]6583 index as calibrated by \citet{Stasinska_2006}:

\begin{equation}
\label{eq:Z_neb}
\log Z_{neb} =
   \log \frac{{\rm (O/H)}}{{\rm ~~(O/H)}_\odot} =
   -0.14 - 0.25 \log {\rm O}_3 {\rm N}_2
\end{equation}

\noindent where we have adopted ${\rm (O/H)}_\odot = 4.9 \times
10^{-4}$ (\citealp*{AllendePrieto_Lambert_Asplund_2001}).

We have chosen to use the O$_3$N$_2$ indicator to estimate the average
oxygen abundance in the ISM of SF galaxies mainly because it is a
single-valued indicator and it can be easily related to the position
of galaxies in the classical BPT diagram. However, this indicator is
affected by the the presence of diffuse ionized gas in galaxies and by
the fact that the N/O ratio depends on the way chemical evolution
proceeded (\citealp*{Chiappini_Romano_Matteucci_2003}). In addition,
for the lowest metallicity galaxies, O$_3$N$_2$ is not sensitive to
O/H anymore, as a wide range of metallicities correspond to the same
value of O$_3$N$_2$, as can be seen in Fig.\ 3 of
\citet{Stasinska_2006}. From that figure, the O/H given by equation
(\ref{eq:Z_neb}) lies towards the upper end of the possible values of
O/H. We considered using the [ArIII]7135/[OIII]5007 (Ar$_3$O$_3$)
index which, as argued by \citet{Stasinska_2006}, does not suffer from
the problems mentioned for the O$_3$N$_2$ index. It turns out that, in
objects where the [ArIII] line could be measured, Ar$_3$O$_3$ and
O$_3$N$_2$ are extremely well correlated (with a Spearman correlation
coefficient of $R_S = 0.58$). Unfortunately, the quality of the SDSS
spectra did not allow us to measure the [ArIII] line intensity with
sufficient accuracy in a large number of objects, and principally in
the zone where it would have been helpful to break the O$_3$N$_2$ vs
O/H degeneracy. Using the \citet{Pilyugin_Thuan_2005} metallicity
calibration based on [OIII]5007/H$\beta$ and [OII]3727/H$\beta$ adds
only a tiny fraction of galaxies.  The same applies when using the O/H
values obtained by \citet{Izotov_etal_2006} from direct methods using
the electron temperature derived from [OIII]4363/[OIII]5007.  When
comparing our measures with theirs, there is a systematic offset of
0.2 dex and a rms of 0.13 dex in $Z_{neb}$ for the 177 objects we have
in common. This effect is greater the lower $Z_{neb}$ is.  We thus
decided to use O$_3$N$_2$ as a nebular metallicity indicator all along
the SF galaxy sequence, keeping in mind that equation (\ref{eq:Z_neb})
will tend to attribute a metallicity $Z_{neb}$ of about 0.2 $Z_\odot$
for the galaxies with the lowest observed O$_3$N$_2$ in our sample.

Other metallicity estimates have been used for galaxies.  For example,
\citet{Tremonti_etal_2004} obtained the nebular metallicities by
comparing the observed line ratios with a large data base of
photoionization models. While a priori appealing, this method is not
devoid of problems, as shown by \citet{Yin_etal_2007}. There is a
systematic offset of -0.28 dex and a rms of 0.09 between our nebular
metallicities and theirs. Their method also yields a larger range of
values for $Z_{neb}$. For the SF sample, their calibration covers from
$Z_{neb} = 0.78$ to 2.70 $Z_\odot$ for the 5 to 95 percentile ranges,
whereas our calibration covers from 0.47 to 1.13 $Z_\odot$ for the
same percentile ranges.

The calibration by \citet{Pettini_Pagel_2004} is more similar to our
own. There is a slight offset of -0.04 dex with respect to our
calibration and the dispersion for the SF sample is 0.03 dex. Their
calibration also stretches a little the $Z_{neb}$ ranges: $Z_{neb} =
0.45$ to 1.36 $Z_\odot$ for the 5 to 95 percentile ranges.

Although we believe that our calibration is likely more reliable, we
have performed all the computations in this paper also with the
\citet{Tremonti_etal_2004} and \citet{Pettini_Pagel_2004}
calibrations. While the results differ in absolute scales, the
qualitative conclusions remain identical.

\subsubsection{$Z_{neb}$ bins}

As seen above, both physical and mathematical motivations make
O$_3$N$_2$ a convenient index to map galaxy positions along the SF
wing in the BPT diagram. From equation (\ref{eq:Z_neb}) one sees that
a given value of $Z_{neb}$ using this index defines a straight line of
unit slope in the BPT diagram.

Our SF sample spans the $Z_{neb} = 0.2$--1.6 $Z_\odot$ range from the
tip of the SF-wing to its bottom. In Fig. \ref{fig:BPT} this
interval is chopped into 6 bins of width $\Delta \log Z_{neb} = 0.13$
dex, except for the one of lowest metallicity which is twice as wide
to include more sources.  Table \ref{tab:ZnebBinsStats} lists some
properties of galaxies in each of these bins, which are hereafter
labeled A--F.  Galaxies inside these bins will be grouped together in
the analysis of star-formation presented in Section
\ref{sec:SFH_results}.  Note that the bias in the determination of
$Z_{neb}$ from O$_3$N$_2$ at low metallicities has no consequence for
our study, since almost all the objects from bin A remain in this bin.

%---------------------------- Table -----------------------------
\begin{table*}
\begin{center}
\begin{tabular}{lcccccc}
\hline
                                           &   Bin A &   Bin B &   Bin C &   Bin D &   Bin E &   Bin F \\
\hline					    
$\log Z_{neb}$ min [Z$_\odot$]             &  -0.710 &  -0.450 &  -0.320 &  -0.190 &  -0.060 &   0.070 \\ 
$\log Z_{neb}$ p05                         &  -0.608 &  -0.435 &  -0.311 &  -0.180 &  -0.053 &   0.071 \\ 
$\log Z_{neb}$ p50                         &  -0.494 &  -0.364 &  -0.242 &  -0.111 &  -0.001 &   0.081 \\ 
$\log Z_{neb}$ p95                         &  -0.454 &  -0.324 &  -0.194 &  -0.064 &   0.054 &   0.112 \\ 
$\log Z_{neb}$ max                         &  -0.450 &  -0.320 &  -0.190 &  -0.060 &   0.070 &   0.200 \\ 
\hline					    
$\log M_\star$ p05 [M$_\odot$]             &   7.146 &   7.938 &   8.693 &   9.277 &   9.800 &  10.089 \\ 
$\log M_\star$ p50                         &   8.319 &   8.922 &   9.452 &   9.958 &  10.460 &  10.678 \\ 
$\log M_\star$ p95                         &   9.313 &   9.634 &  10.083 &  10.646 &  11.073 &  11.154 \\ 
\hline					    
$\langle \log t_\star \rangle_L$ p05 [yr]  &   6.755 &   7.428 &   7.695 &   7.915 &   8.127 &   8.167 \\ 
$\langle \log t_\star \rangle_L$ p50       &   7.762 &   8.155 &   8.385 &   8.567 &   8.726 &   8.724 \\ 
$\langle \log t_\star \rangle_L$ p95       &   8.556 &   8.900 &   9.139 &   9.255 &   9.276 &   9.255 \\ 
\hline					    
$\log L_{H\alpha}$ p05 [L$_\odot$]         &   5.167 &   5.139 &   5.492 &   5.988 &   6.551 &   6.956 \\ 
$\log L_{H\alpha}$ p50                     &   6.475 &   6.475 &   6.750 &   7.149 &   7.544 &   7.816 \\ 
$\log L_{H\alpha}$ p95                     &   7.834 &   7.736 &   7.961 &   8.172 &   8.426 &   8.559 \\ 
\hline					    
$\log b$ p05                               &  -0.157 &  -0.392 &  -0.698 &  -0.947 &  -0.973 &  -0.890 \\ 
$\log b$ p50                               &   0.769 &   0.436 &   0.211 &  -0.006 &  -0.213 &  -0.199 \\ 
$\log b$ p95                               &   1.527 &   1.192 &   0.905 &   0.644 &   0.366 &   0.272 \\ 
\hline
\end{tabular}
\end{center}
\caption{Statistics of properties in bins A--F.}
\label{tab:ZnebBinsStats}
\end{table*}
%---------------------------- Table -----------------------------

\subsection{Extinctions}
\label{sec:AV_neb}

\subsubsection{Stellar extinction}

As explained in SEAGal I, \starlight\ also returns an estimate of the
stellar visual extinction, $A_V^\star$, modeled as due to a foreground
dust screen. This is obviously a simplification of a complex problem
(\citealp*{Witt_Thronson_Capuano_1992}), so that $A_V$ should be called
a dust attenuation parameter instead of extinction, although we do not
make this distinction.  Previous papers in this series have used the
\citet*[CCM]{Cardelli_Clayton_Mathis_1989} reddening law, with $R_V =
3.1$.  In order to probe different recipes for dust attenuation, we
have selected 1000 galaxies at random from the SF$^{hq}$ sample and
fitted them with four other functions: the starburst attenuation law
of \citet*{Calzetti_Kinney_Storchi-Bergmann_1994}, the SMC and LMC
curves from \citet{Gordon_etal_2003} and the $\lambda^{-0.7}$ law used
by \citet{Kauffmann_etal_2003a}.

We find that the quality of the spectral fits remains practically
unchanged with any of these 5 laws. Averaging over all galaxies the
SMC law yields slightly better $\chi^2$'s, followed closely by the
Calzetti, $\lambda^{-0.7}$, LMC, and CCM, in this order. As expected,
these differences increase with the amount of dust, as measured by the
derived $A_V^\star$ values or by H$\alpha$/H$\beta$.  Yet, KS-tests
showed that in no case the distributions of $\chi^2$'s differ
significantly.  This implies that the choice of reddening law {\em
cannot} be made on the basis of fit quality. A wider spectral coverage
would be needed for a definitive empirical test.

When using different recipes for the dust attenuation, the synthesis
algorithm has to make up for the small variations from one curve to
another by changing the population vector and the value of
$A_V^\star$.  To quantify these changes we compare results obtained
with the Calzetti and CCM curves, and consider only the most extincted
objects. Compared to the results for a CCM law, with the Calzetti law
the mean stellar age decreases by 0.09 dex on the median
(qualitatively in agreement with the results reported in Fig 6 of
\citealp{Panter_etal_2007_Moped}), the mean stellar metallicity
increases by 0.05 dex, $A_V^\star$ increases by 0.07 mag and stellar
masses decrease by 0.02 dex.  These differences, which are already
small, should be considered upper limits, since they are derived from
the most extincted objects. Somewhat larger differences are found when
using the SMC and $\lambda^{-0.7}$ laws. For instance, compared to the
Calzetti law, the SMC law produces mean stellar ages 0.15 dex younger
and masses 0.07 dex smaller, again for the most extincted objects.

We have opted to use the Calzetti law in our \starlight\ fits and
emission line analysis. The reasons for this choice are twofold: (1)
The Calzetti law yields physical properties intermediate between the
SMC and CCM laws, and (2) this law was built up on the basis of
integrated observations of SF galaxies, similar to the ones studied in
this paper. In any case, the experiments reported above show that this
choice has little impact upon the results.

\starlight\ also allows for population dependent extinctions. Although
tests with these same 1000 galaxies sample show that in general one
obtains larger $A_V^\star$ for young populations, as expected,
simulations show that, as also expected, this more realistic modeling
of dust effects is plagued by degeneracies which render the results
unreliable (see also \citealp{Panter_etal_2007_Moped}; SEAGal V). We
therefore stick to our simpler but more robust single $A_V^\star$
model.

\subsubsection{Nebular extinction}
\label{sec:av_neb}

The nebular V-band extinction was computed from the H$\alpha$/H$\beta$
ratio assuming a \citet{Calzetti_Kinney_Storchi-Bergmann_1994} law:

\begin{equation}
\label{eq:AV_neb}
A_V^{neb} = 7.96 \log {\rm
   \frac{(H\alpha/H\beta)_{obs}}{(H\alpha/H\beta)_{int}} }
\end{equation}

\ni where ${\rm (H\alpha/H\beta)_{obs}}$ and ${\rm
(H\alpha/H\beta)_{int}}$ are the observed and intrinsic ratio
respectively. Instead of assuming a constant value, we take into
account the metallicity dependence of ${\rm (H\alpha/H\beta)_{int}}$,
which varies between 2.80 and 2.99 for $Z_{neb}$ in the 0.1 to 2.5
$Z_\odot$ range, as found from the photoionization models in SEAGal
III\footnote {Note that the models take into account collisional
excitation of Balmer lines, so that at low metallicities the intrinsic
H$\alpha$/H$\beta$ is different from the pure case B recombination
value.}.

We obtain the intrinsic ratio as follows.  We start by assuming ${\rm
(H\alpha/H\beta)_{int}} = 2.86$, from which we derive a first guess
for $A_V^{neb}$.  We then use the dereddened ${\rm [OIII]}\lambda5007$
and ${\rm [NII]}\lambda6584$ line fluxes to calculate $Z_{neb}$
(eq.~\ref{eq:Z_neb}).  From our sequence of photoionization models
(SEAGal III) and $Z_{neb}$, we derive a new estimate for ${\rm
(H\alpha/H\beta)_{int}}$, and hence $A_V^{neb}$
(eq.~\ref{eq:AV_neb}). It takes a few iterations (typically 2--3) to
converge.

For 1.6\% of the objects (1.1\% for the SF$^{hq}$ sample),
H$\alpha$/H$\beta$ is smaller than the intrinsic value, which leads to
$A_V^{neb} < 0$. In such cases, we assume $A_V^{neb} = 0$.  We have
corrected both the [OIII]/H$\beta$ and [NII]/H$\alpha$ line ratios for
dust attenuation for the remainder of our analysis.

%---------------------------- Figure ----------------------------
\begin{figure}
   \includegraphics[width=0.5\textwidth]{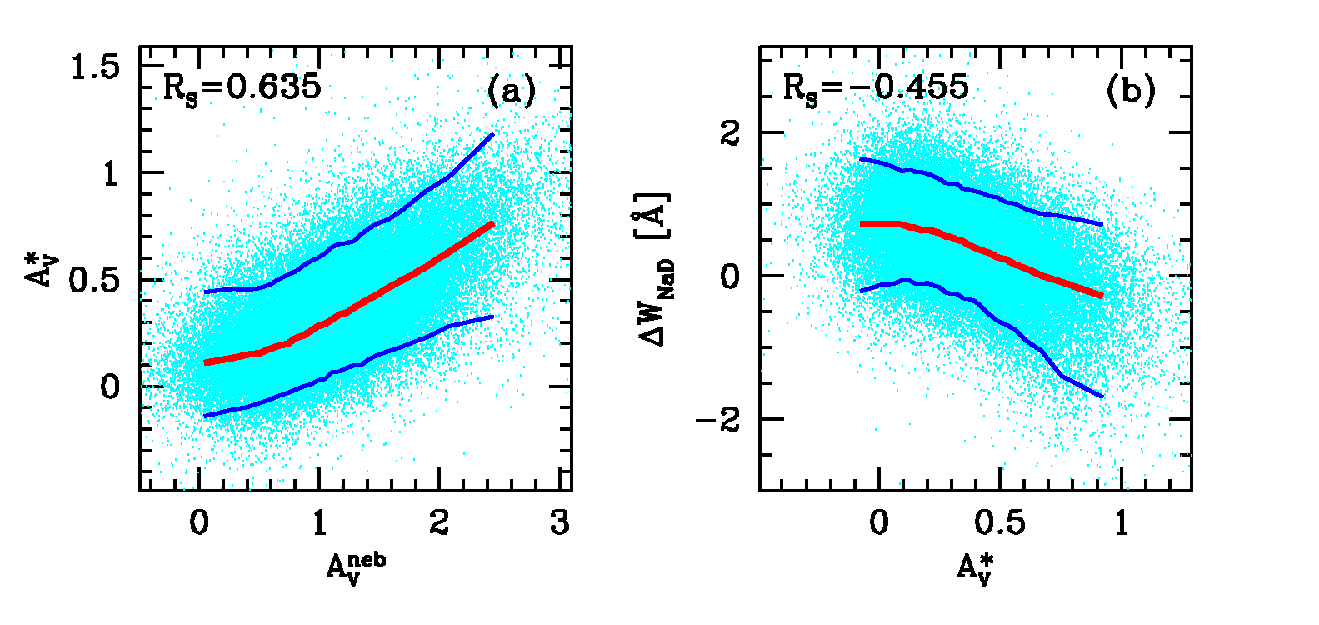}
   \caption{(a) Relation between stellar and nebular extinctions for
the SF sample. Lines indicate the 5, 50 and 95 percentiles. (b)
Equivalent width of the ISM component of the NaD doublet (as measured
from residual spectra) against the stellar extinction.}
\label{fig:ResidualsNaD}
\end{figure}
%---------------------------- Figure ----------------------------

We find that $A_V^{neb}$ and $A_V^\star$ are strongly correlated, as
shown in Fig \ref{fig:ResidualsNaD}a. A robust linear fit including
all points yields $A_V^{neb} = 0.34 + 2.28 A_V^\star$.  The ionized
gas thus suffers $\sim$ twice as much extinction as the stellar
continuum, corroborating the results reported in
\citet{Stasinska_etal_2004} (with a different methodology) and SEAGal
I (obtained with a smaller sample, different version of \starlight\ and
a CCM extinction curve), and in agreement with detailed studies of
nearby SF-galaxies (\citealp{Calzetti_Kinney_Storchi-Bergmann_1994}).
We also find that the difference between nebular and stellar
extinctions increases systematically as the mean age of the stellar
population increases.

Given the spatial association of the line emitting gas and the
ionizing populations, these results ultimately imply a breakdown of
our simple single-$A_V^\star$ modelling. In fact, \starlight\
experiments with population dependent $A_V^\star$ point in the same
direction, i.e, the need to allow young populations to suffer more
extinction than older ones. To evaluate to which extent this
simplification affects the results reported in this paper, Sec.\
\ref{sec:samples_YAV} presents experiments where the extinction of
$t_\star \le 10$ Myr components is set according to the empirical
relation $A_V^{neb}(A_V^\star)$ found above.

\subsubsection{Interstellar absorption as traced by the NaD doublet}

The most conspicuous spectroscopic feature of the cold ISM in the
optical range is the NaD doublet at $\lambda\lambda$5890,5896 \AA.
For a constant gas to dust ratio, the strength of this feature, which
measures the amount of cold gas in front of the stars, should
correlate with $A_V^\star$, as found for far-IR bright starburst
galaxies (\citealp{Heckman_etal_2000}). To perform this test for our
sample, we first measure the flux of the NaD doublet in the residual
spectrum, integrating from 5883 to 5903 \AA. We thus remove the
stellar component of this feature, which is also present in stellar
atmospheres, particularly late type stars
(\citealp*{Jacoby_Hunter_Christian_1984}; \citealp{Bica_etal_1991}).
In principle, this is a more precise procedure than estimating the
stellar NaD from its relation to other stellar absorption lines
(\citealp{Heckman_etal_2000, Schwartz_Martin_2004}), but since the NaD
window was masked in all fits (precisely because of its possible
contamination by ISM absorption), the stellar NaD predicted by the
fits rely entirely on other wavelengths, so in practice this is also
an approximate correction.  The residual flux is then divided by the
continuum in this range (defined as the median synthetic flux in the
5800--5880 plus 5906-5986 \AA\ windows), yielding the excess
equivalent width $\Delta W_{\rm NaD}$, which says how much stronger
(more negative) the NaD feature is in the data with respect to the
models.

Fig \ref{fig:ResidualsNaD}b shows the relation between $\Delta W_{\rm
NaD}$ and $A_V^\star$.  The plot shows that these two independently
derived quantities correlate strongly.  Intriguingly, $\Delta W_{\rm
NaD}$ converges to $0.8$ \AA\ in the median as $A_V^\star
\rightarrow 0$. We interpret this offset from $\Delta W_{\rm NaD} = 0$
as due to the fact that the stars in the STELIB library have a
Galactic ISM component in their NaD lines. This propagates to our
spectral models, leading to an overprediction of the stellar NaD
strength, and thus $\Delta W_{\rm NaD} > 0$ when the ISM absorption
approaches zero.

Regardless of such details, the discovery of this astrophysically
expected correlation strengthens the confidence in our
analysis. Furthermore, it opens the interesting prospect of measuring
the gas-to-dust ratio and study its relation with all other galaxy
properties at hand, from nebular metallicities to SFHs.  This goes
beyond the scope of the present paper, so we defer a detailed analysis
to a future communication.

%***************************************************************%
%                                                               %
%                      Correlations                             %
%                                                               %
%***************************************************************%

\section{Correlations with nebular metallicity}
\label{sec:Correlations}

Galaxy properties change substantially from the tip of the SF-wing,
where small, metal-poor HII-galaxy-like objects live, to its bottom,
populated by massive, luminous galaxies with large bulge-to-disk
ratios and rich in metals (\citealp{Kennicutt_1998}). The simplest way
to investigate these systematic trends is to correlate various
properties with the nebular metallicity (e.g.,
\citealp{Tremonti_etal_2004, Brinchmann_etal_2004}).

In this section we correlate $Z_{neb}$ with both observed and physical
properties extracted from our stellar population fits. This
traditional analysis, based on current or time-averaged properties,
helps the interpretation of the more detailed study of time-dependent
SFHs presented in the next sections.  In fact, this is the single
purpose of this section. Since most of the results reported in this
section are already know or indirectly deducible from previous work,
we will just skim through these correlations.

%---------------------------- Figure ----------------------------
\begin{figure*}
   \includegraphics[width=\textwidth]{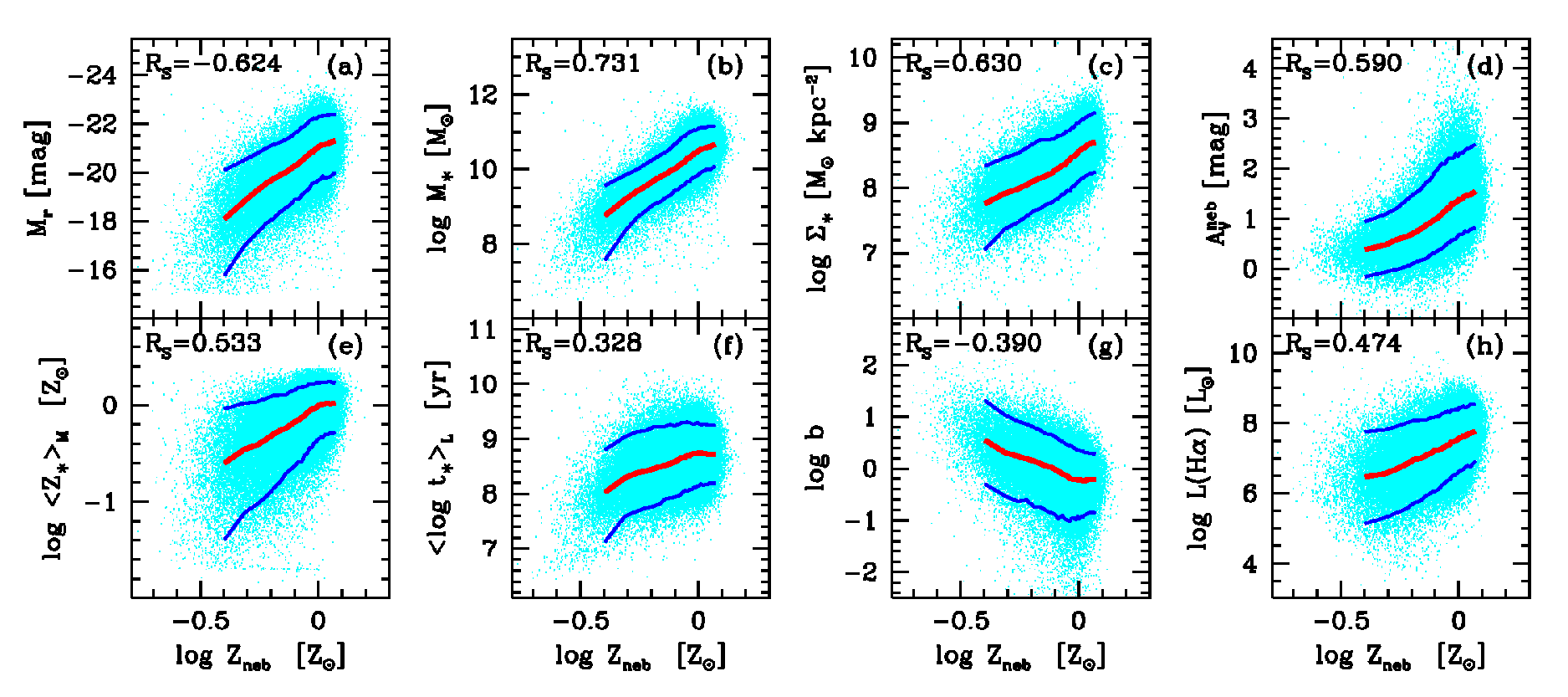}
   \caption{Correlations of $Z_{neb}$ and (a) the absolute r-band
   magnitude, (b) the stellar mass, (c) the surface mass density, (d)
   the nebular extinction, (e) the mean stellar metallicity, (f) the
   mean stellar age, (g) the ratio of current to mean past SFR, and (h)
   the H$\alpha$ luminosity.  Numbers in each panel report the
   Spearman rank correlation coefficient, and the lines mark the 5, 50
   and 95\% percentiles of 25 bins, 3292 points in each bin. The right
   hand scale in panel (h) is also $\log L_{H\alpha}$, but in units of
   $2 \times 10^8 L_\odot$, such that it can also be read as an
   estimate of the current SFR in $M_\odot\,$yr$^{-1}$ (Section
   \ref{sec:SFRxHa}).}
\label{fig:ZnebCorrelations}
\end{figure*}
%---------------------------- Figure ----------------------------

Fig \ref{fig:ZnebCorrelations}a shows $Z_{neb}$ against absolute
r-band magnitude. This is the luminosity-nebular metallicity relation,
previously studied by many authors, and interpreted in terms of a
mass-metallicity relation. Fig \ref{fig:ZnebCorrelations}b shows our
version of the $M_\star$--$Z_{neb}$ relation.  Because of the expected
bias in our $Z_{neb}$ estimate at the lowest metallicities (see
Section \ref{sec:Z_neb}), we expect the real mass-metallicity relation
to be flatter at low $M_\star$ than seen in this plot.

As shown by \citet{Kauffmann_etal_2003b}, stellar mass and stellar
surface mass density ($\Sigma_\star$) are very strongly related. It is
thus no surprise to find that $\Sigma_\star$ also correlates with
$Z_{neb}$, as shown in Fig \ref{fig:ZnebCorrelations}c. Our definition
of $\Sigma_\star$ is the same as adopted by \citet{Kauffmann_etal_2003b},
namely $\Sigma_\star = M_\star / 2\pi r_{50,z}^2$, where $r_{50,z}$ is
the half light Petrosian radius in the z-band.

Fig \ref{fig:ZnebCorrelations}d shows how nebular extinction increases
systematically with $Z_{neb}$. One factor which surely contributes to
this relation is the rise in dust grain formation with increasing gas
metallicity, but other factors may come into play as well
(\citealp{Stasinska_etal_2004}).

Fig \ref{fig:ZnebCorrelations}e shows how nebular and stellar
metallicities correlate.  This important relation, first presented in
SEAGal I (for different sample, SSP base and $Z_{neb}$ scale), shows
that stellar and ISM chemical enrichment levels scale with each other,
as one would expect on the basis of simple chemical evolution
scenarios. The large scatter in Fig \ref{fig:ZnebCorrelations}e is
mostly intrinsic (as we verified comparing the relation obtained for
data of different qualities), in qualitative agreement with the idea
that stellar and nebular metallicities reflect different evolutionary
phases and react differently to the several processes which regulate
the chemical evolution of galaxies. A similar relation was obtained by
\citet{Gallazzi_etal_2005} using different methods to estimate both
stellar and nebular abundances.  Even though we express both
quantities in solar units, these two metallicities are derived by such
radically different means that, as discussed in SEAGal V, they should
not be compared in quantitative terms.

The relation between the mean stellar age $\langle \log t_\star
\rangle_L$ and $Z_{neb}$, shown in Fig \ref{fig:ZnebCorrelations}f,
reflects the fact that young stars have a larger share of the light
output at the tip of the SF wing than at its bottom, where old
populations have a greater weight. Metal-rich SF galaxies thus have a
more continuous star-forming history than metal-poor ones, which are
often dominated (in light, but not mass) by the latest generation of
stars (e.g., \citealp{Corbin_etal_2006}).  This is another way to look
at metallicity--age trend, discussed previously in the analysis of
Fig.~\ref{fig:STARLIGHT_fits}. Ultimately, this relation represents a
summary of chemical evolution, in the sense that more evolved systems
have a more enriched ISM.  In a related vein, Fig
\ref{fig:ZnebCorrelations}g shows how the ratio of current to mean
past SFR (defined in Section \ref{sec:bScalo_def}) varies along the
metallicity sequence of SF galaxies. This indicates that the
lower-metallicity galaxies are slower in forming stars. When one
considers the mass-metallicity relation (Fig
\ref{fig:ZnebCorrelations}b), this is just another way of looking at
the downsizing effect (\citealp{Heavens_etal_2004_Moped,
Thomas_etal_2005, Mateus_etal_2006_SEAGal2}).  Finally, Fig
\ref{fig:ZnebCorrelations}h shows the relation between reddening
corrected H$\alpha$ luminosity and $Z_{neb}$. The y-axis is given in
units such that the values correspond $\sim$ to the current SFR in
$M_\odot\,$yr$^{-1}$ (see Section \ref{sec:SFRxHa}). The correlation,
although statistically unquestionable, has a large scatter. This
implies that galaxies in the 6 $Z_{neb}$ bins defined in Fig
\ref{fig:BPT} have heavily overlapping current SFRs.  Section
\ref{sec:SFH_results} presents independent confirmation of this fact.

As expected, all correlations discussed above are also present for the
SF$^{hq}$ sub-sample. For most they are in fact somewhat stronger,
whereas for samples defined with less stringent criteria the
correlation strengths weaken, indicating that noise in the data is
responsible for part of the scatter in these relations.  Finally, as is
widely known and can be deduced from Fig \ref{fig:ZnebCorrelations}
itself, there are many inter-relations between galaxy properties. Our
use of $Z_{neb}$ as the ``independent'' variable axis in Fig
\ref{fig:ZnebCorrelations} is not meant to indicate that $Z_{neb}$ is
the underlying cause of the correlations; it simply reflects our
interest in mapping physical properties of galaxies along the SF wing
of the seagull in the BPT diagram.

%***************************************************************%
%                                                               %
%                   Star Formation Histories                    %
%                                                               %
%***************************************************************%

\section{Methods to investigate star formation histories}
\label{sec:SFH_theory}

The main goal of this paper is to study how the SFH varies among SF
galaxies.  Most other investigations in this same line used
absorption, emission or continuum spectral indices such as the 4000
\AA\ break, the H$\delta$ absorption, the K, G and Mg bands, or the
H$\alpha$ luminosity and equivalent widths to characterize the SFH
(e.g., \citealp{Raimann_etal_2000, Kong_etal_2003,
Kauffmann_etal_2003b, CidFernandes_Leao_Lacerda_2003,
Brinchmann_etal_2004, Westera_etal_2004}). Our approach, instead, is
to infer the SFH from detailed pixel-by-pixel fits to the full
observed spectrum, thus incorporating all available information.

Whereas our previous work concentrated on the first moments of the age
and metallicity distributions, here we present some basic formalism
towards a robust description of SFHs as a function of time.  From the
point of view of methodology, these may be regarded as
``second-order'' products. Astrophysically, however, recovering the
SFH of galaxies is of prime importance.  SEAGal V presented our first
results in this direction, including empirically derived
time-dependent mean stellar metallicities. In this section we expand
upon these results, exploring new ways to handle the output of the
synthesis, focusing of the SFHs.

\subsection{Compression methods}
\label{sec:STARLIGHT}

As reviewed in Section \ref{sec:STARLIGHTfits}, \starlight\ decomposes
an observed spectrum in terms of a sum of SSPs, estimating the $x_j$
($j = 1\cdots N_\star$) fractional contribution of each population to
the flux at $\lambda_0 = 4020$ \AA. For this work we used a base of
$N_\star = 150$ SSPs from BC03, spanning 25 ages between $t_{\star,j}
= 1$ Myr and 18 Gyr, and 6 metallicities ($0.005 \le Z_{\star,j} \le
2.5 Z_\odot$).  Example fits were shown in Fig
\ref{fig:STARLIGHT_fits}.

Not surprisingly, the 150 components of the population vector
($\vec{x}$) are highly degenerate due to noise, and astrophysical plus
mathematical degeneracies, as confirmed by simulations in
\citet{CidFernandes_etal_2004} and SEAGal I.  These same simulations,
however, proved that {\it compressed} versions of the population
vector are well recovered by the method.

Different compression approaches exist among spectral synthesis codes.
In MOPED (\citealp*{Heavens_Jimenez_Lavah_2000_Moped};
\citealp{Reichardt_etal_2001_Moped, Panter_etal_2003_Moped,
Panter_etal_2007_Moped, Heavens_etal_2004_Moped}), for instance,
compression is done {\it a priori}, replacing the full spectrum by a
set of numbers associated to each of the $N_\star + 1$ parameters (the
mass fractions and metallicities in several time bins plus a dust
parameter).  STECMAP (\citealp{Ocvirk_etal_2006}) performs a
compression by requiring the resulting SFH to be relatively smooth.
The preference for a smooth solution over a ragged one is effectively
a prior, but the algorithm adjusts the degree of smoothing in a data
driven fashion, so we may call it an ``on the fly'' compression
method.  The same can be said about VESPA
(\citealp{Tojeiro_etal_2007_Vespa}), a new code which combines
elements from these two approaches.  \starlight\ is less sophisticated
in this respect. Its only built-in compression scheme is that the
final stages of the fit (after the Markov Chains reach convergence)
are performed with a reduced base comprising the subset of the
original $N_\star$ populations which account for $\ge 99\%$ of the
light.  For our parent sample of 573141 galaxies the average size of
this subset is $\overline{N_\star^{\rm eff}} = 24$ populations, while
for the 82302 galaxies in the SF sample $\overline{N_\star^{\rm eff}}
= 41$. (This difference happens because the full sample has many old,
passive systems, which require relatively few SSPs, whereas SF
galaxies have more continuous SF regimes, thus requiring more SSPs to
be adequately fit.)  Compression beyond this level must be carried out
{\it a posteriori} by the user. As explained in the next section, in
this study we in fact compress this information into only four age
bins by smoothing the population vectors.

Previous papers in this series have taken this {\it a posteriori}
compression approach to its limit, condensing the whole age
distribution to a single number, the mean stellar age:

\begin{equation}
\label{eq:logt_ave}
\langle \log t_\star \rangle_L =
    \sum_{j=1}^{N_\star} x_j \log t_{\star,j}
\end{equation}

\ni where the subscript $L$ denotes a light-weighted average.
Mass-weighted averages are readily obtained replacing $\vec{x}$ by the
mass-fraction vector $\vec{\mu}$. Similarly, stellar metallicities
were only studied in terms of their mass-weighted mean value:

\begin{equation}
\label{eq:Z_ave}
\langle Z_\star \rangle_M =
    \sum_{j=1}^{N_\star} \mu_j Z_{\star,j}
\end{equation}

Simulations show that both of these quantities have small
uncertainties and essentially no bias. Regarding practical
applications, these first moments proved useful in the study of
several astrophysical relations, some of which have just been
presented in Section \ref{sec:Correlations} (see
Fig.~\ref{fig:ZnebCorrelations}).  Notwithstanding their simplicity,
robustness and usefulness, these averages throw away all time
dependent information contained in the population vector, thus
hindering more detailed studies of galaxy evolution. In what follows
we explore novel ways to deal with the population vector which
circumvent this limitation.

\subsection{Star Formation Rate as a Function of Time}

One alternative to characterize higher moments of the SFH is to bin
$\vec{x}$ onto age-groups, a strategy that goes back to
\citeauthor{Bica_1988} (\citeyear{Bica_1988}; see also
\citealp{Schmidt_etal_1991, CidFernandes_etal_2001}). Though useful,
this approach introduces the need to define bin-limits, and produces a
discontinuous description of the SFH.

A method which circumvents these disadvantages is to work with a {\em
smoothed} version of the population vector. We do this by applying a
gaussian filter in $\log t_\star$, with a FWHM of 1 dex. Given that
our base spans $\sim 4$ orders of magnitude in $t_\star$, this heavy
smoothing is equivalent to a description in terms of $\sim 4$ age
groups, but with the advantage that $\vec{x}_s$ can be sampled
continuously in $\log t_\star$.  This approach is analogous to
smoothing a noisy high-resolution spectrum to one of lower resolution,
but whose large-scale features (colours, in this analogy) are more
robust. From the results in SEAGal I, where it was shown that 3 age
groups are reliably recovered, we expect this smoothing strategy to be
a robust one. Furthermore, averaging over a large number of objects
minimizes the effects of uncertainties in the smoothed SFH for
individual galaxies.

Technically, the issue of age resolution in population synthesis is a
complex one. \citet{Ocvirk_etal_2006}, for instance, find that bursts
must be separated by about 0.8 dex in $\log t_\star$ to be well
distinguished from one another with their SFH inversion
method. Considering that the age range spanned by our base is 4.2 dex
wide, one obtains 5 ``independent'' time bins. This number is similar
to that (6 bins) used by \citet*{Mathis_Charlot_Brinchmann_2006} to
describe the SFH of SDSS galaxies (covering a wider $\lambda$-range
than those simulated by \citeauthor{Ocvirk_etal_2006} but at lower
S/N) with a variant of the MOPED code. Up to 12 time bins were used in
other applications of MOPED.  \citet{Panter_etal_2007_Moped} argue
that this may be a little too ambitious for individual galaxies, but
uncertainties in this overparameterized description average out in
applications to large samples.  \citet{Tojeiro_etal_2007_Vespa} have a
useful discussion on the number of parameters that can be recovered
with synthesis methods. By using their VESPA code and calculating the
number of parameters on the fly for each individual object, they find
that tipically 2--8 parameters can be robustely recovered for SDSS
spectra. Hence, despite the complexity of the issue, there seems to be
some general consensus that the age resolution which can be achieved
in practice is somewhere between 0.5 and 1 dex, so our choice of
smoothing length is clearly on the conservative side.

A further advantage of this continuous description of the SFH is that
it allows a straight-forward derivation of a star-formation rate
(SFR). Recall that we describe a galaxy's evolution in terms of a
succession of instantaneous bursts, so a SFR is not technically
definable unless one associates a duration to each burst.  The ${\rm
SFR}(t_\star)$ function is constructed by sampling the smoothed
mass-fraction vector $\vec{\mu}^c_s$ in a quasi-continuous grid from
$\log t_\star = 5.6$ to 10.5 in steps of $\Delta \log t_\star = 0.1$
dex, and doing

\begin{equation}
\label{eq:SFR}
       {\rm SFR}(t_\star) = \frac {d M^c_\star(t_\star)}{dt_\star} \approx
       \frac{\Delta M^c_\star(t_\star)}{\Delta t_\star} =
       \frac{M_\star^c \log e}{t_\star}\frac {\mu^c_s(t_\star)}{\Delta 
\log t_\star}
\end{equation}

\ni where $M_\star^c$ is the total mass {\em converted to stars} over
the galaxy history until $t_\star = 0$, and $\vec{\mu}^c_s(t_\star)$
is the fraction of this mass in the $t_\star$ bin.\footnote{The
superscript $c$ is introduced to distinguish $M_\star^c$ from the
mass still locked inside stars ($M_\star$), which must be corrected
for the mass returned to the ISM by stellar evolution.  This
distinction was not necessary in previous SEAGal papers, which dealt
exclusively with $M_\star$ and its associated mass-fraction vector
($\vec{\mu}$). When computing SFRs, however, this difference must be
taken into account.  From the BC03 models for a \citet{Chabrier_2003}
IMF, a $10^{10}$ yr old population has $M_\star^c \sim 2 M_\star$,
i.e., only half of its initial mass remains inside stars nowadays.}

We can also define the time dependent {\em specific} SFR:

\begin{equation}
\label{eq:SSFR}
       {\rm SSFR}(t_\star) =
       \frac {1}{M^c_\star}
       \frac {d M^c_\star(t_\star)}{dt_\star} \approx
       \frac{\log e}{t_\star}\frac {\mu^c_s(t_\star)}{\Delta \log t_\star}
\end{equation}

\ni which measures the pace at which star-formation proceeds with
respect to the mass already converted into stars. This is a better
quantity to use when averaging the SFH over many objects, since it
removes the absolute mass scale dependence of equation (\ref{eq:SFR}).

Three clarifying remarks are in order. (1) All equations above are
marginalized over $Z_\star$, i.e., ${\rm SFR}(t_\star) =
\sum_{Z_\star}{\rm SFR}(t_\star, Z_\star)$ measures the rate at which
gas is turned into stars of {\em any} metallicity. (2) The upper age
limit of our base (18 Gyr) is inconsistent with our adopted cosmology,
which implies an 13.5 Gyr Universe. Given the uncertainties in stellar
evolution, cosmology, observations and in the fits themselves, this is
a merely formal inconsistency, and, in any case, components older than
13.5 Gyr can always be rebinned to a cosmologically consistent time
grid if needed.  (3) Finally, since our main goal is to compare the
{\em intrinsic} evolution of galaxies in different parts of the SF
wing in the BPT diagram, throughout this paper we will consider ages
and lookback times in the context of stellar-evolution alone. In other
words we will {\em not} translate $t_\star$ to a cosmological lookback
time frame, which would require adjusting the $t_\star$ scale by
adding the $z$-dependent lookback time of each galaxy.

\subsection{Mass Assembly Histories}

Another way to look at the population vector is to compute the total
mass converted into stars as a function of time:

\begin{equation}
\label{eq:MAH}
\eta^c_\star(t_\star) =
     \sum_{t_{\star,j} > t_\star} \mu^c_j
\end{equation}

\ni which is a cumulative function that grows from 0 to 1, starting at
the largest $t_\star$, tracking what fraction of $M^c_\star$ was
converted to stars up to a given lookback time.

We sample $\eta^c_\star$ in the same $\log t_\star = 5.6$--10.5 grid
used to describe the evolution of the SFR, but here we operate on
the original population vector, not the smoothed one. Since most of
the mass assembly happens at large $t_\star$, computing $\eta^c_\star$
with the smoothed SFHs leads to too much loss of resolution. In
essence, however, $\eta^c_\star(t_\star)$ and ${\rm
SSFR}_\star(t_\star)$ convey the same physical information in
different forms.

%***************************************************************%
%                                                               %
%                            Halpha                             %
%                                                               %
%***************************************************************%

\section{The current SFR}
\label{sec:SFRxHa}

The most widely employed method to measure the ``current'' SFR is by
means of the H$\alpha$ luminosity (\citealp{Kennicutt_1983,
Kennicutt_1998, Hopkins_etal_2003}). We have just devised ways of
measuring the time dependent SFR which rely exclusively on the stellar
light, from which one can define a current SFR averaging over a
suitably defined time interval.  Before proceeding to the application
of these tools to study the detailed SFHs of galaxies, this section
compares these two independent methods to estimate the current SFR.

The purpose of this exercise is three-fold. First, it serves as yet
another sanity check on the results of the synthesis. Secondly, it
allows us to define in an objective way the ratio of current to
past-average SFR, often referred to as Scalo's $b$ parameter
(\citealp{Scalo_1986}), which is a useful way to summarize the SFH of
galaxies (e.g., \citealp{Sandage_1986, Brinchmann_etal_2004}). Finally,
defining and calibrating a synthesis-based measure of current SFR
equivalent to that obtained with H$\alpha$, allows one to estimate the
current SFR in galaxies where H$\alpha$ is {\em not} powered
exclusively by young stars. This turns out to be very useful in
studies of AGN hosts (\citealp[in prep.]{Torres-Papaqui_etal_2007}).

\subsection{Current SFR from H$\alpha$ luminosity}

For a SFR which is constant over times-scales of the order of the
lifetime of massive ionizing stars ($t_{ion} \sim 10$ Myr), the rate
of H-ionizing photons converges to

\begin{equation}
\label{eq:QH}
Q_H = {\rm SFR} \, {\cal N}_H({\rm IMF},Z_\star)
\end{equation}

\ni where ${\cal N}_H$ is the number of $h\nu > 13.6$ eV photons
produced by a SSP of unit mass over its life (in practice, over 95\%
of the ionizing radiation is produced in the first 10 Myr of
evolution). We computed ${\cal N}_H$ by integrating the $Q_H(t)$
curves for SSPs using the tables provided by BC03, obtaining ${\cal
N}_H = 9.12$, 7.08, 6.17, 5.62, 4.47 and $3.16 \times 10^{60}$
photons$\,$M$_\odot^{-1}$ for $Z_\star = 0.005$, 0.02, 0.2, 0.4, 1 and
2.5 $Z_\odot$, respectively, for a \citet{Chabrier_2003} IMF between
0.1 and 100 M$_\odot$.\footnote{${\cal N}_H$ is 1.66 times smaller for
a Salpeter IMF within the same mass limits.}

One in every 2.226 ionizing photons results in emission of an
H$\alpha$ photon, almost independently of nebular conditions
(\citealp{Osterbrock_Ferland_2006}).  This assumes Case B
recombination and that no ionizing photon escapes the HII region nor
is absorbed by dust. Adopting the Chabrier IMF and the $Z_\odot$ value
of ${\cal N}_H$ leads to:

\begin{equation}
\label{eq:SFR_LHa}
{\rm SFR}_{H\alpha}
   = \frac{2.226 L_{H\alpha}}{{\cal N}_H h\nu_{H\alpha}}
   = 2 M_\odot yr^{-1} \left(
  \frac{L_{H\alpha}}{10^8 L_\odot} \right)
\end{equation}

This calibration is strongly dependent on the assumed IMF and
upper stellar mass limit. Given its reliance on the most massive
stars, which comprise a tiny fraction of all the stars formed in a
galaxy, SFR$_{H\alpha}$ involves a large IMF-dependent extrapolation,
and thus should be used with care.

\subsection{Current SFR from the spectral synthesis}

The SFR from spectral synthesis is based on all stars that contribute
to the visible light, and thus should be more representative of the
true SFR. We define a mean ``current'' SFR from our time dependent
SFHs using equation (\ref{eq:MAH}) to compute the mass converted into
stars in the last $\tau$ years, such that

\begin{equation}
\label{eq:SFR_synthesis}
\overline{{\rm SFR}_\star}(\tau) =
M_\star^c \frac{1 - \eta^c_\star(\tau)}{\tau}
\end{equation}

\ni is the mean SFR over this period. Because of the discrete nature
of our base, the function $\overline{{\rm SFR}_\star}(\tau)$ has a
``saw-tooth'' appearance, jumping every time $\tau$ crosses one of the
$t_{\star,j}$'s bin borders.

For the reasons discussed in Section \ref{sec:STARLIGHT}, it is
desirable to include components spanning $\sim 1$ dex in age to obtain
robust results. Since our base starts at 1 Myr, $\tau \sim 10$ Myr
would be a reasonable choice.  This coincides with the minimum
time-scale to obtain $\overline{{\rm SFR}_\star}(\tau)$ estimates
comparable to those derived from $L_{H\alpha}$, which are built upon
the assumption of constant SFR over $\tau \ge t_{ion} \sim 10$
Myr. Our base ages in this range are $t_{\star,j} = 10$, 14, 25, 40
and 55 Myr.

\subsection{Synthesis versus H$\alpha$-based current SFRs}

\label{sec:bScalo_def}

%---------------------------- Figure ----------------------------
\begin{figure*}
   \includegraphics[width=\textwidth]{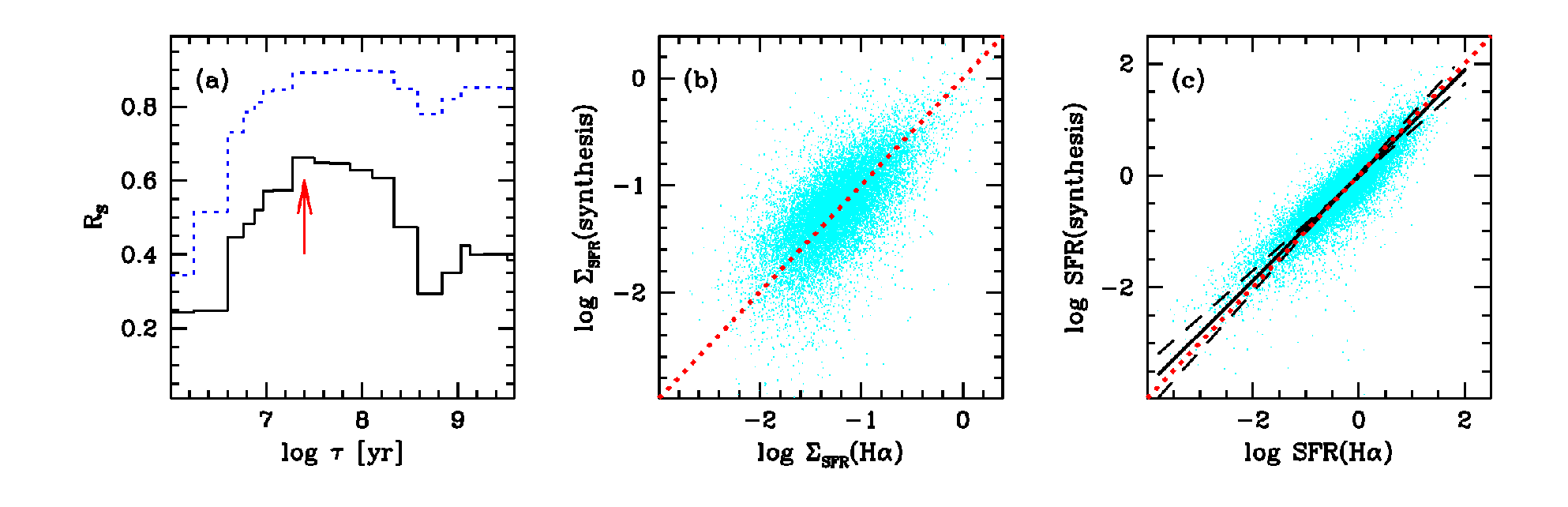}
\caption{(a) The solid line shows the Spearman coefficient ($R_S$) of
   the $\Sigma_{SFR}({\rm synthesis}) \times \Sigma_{SFR}(H\alpha)$
   correlation for different values of $\tau$ in equation
   (\ref{eq:SFR_synthesis}). The dotted line indicates the strength of
   the ${\rm SFR}({\rm synthesis}) \times {\rm SFR}(H\alpha)$
   correlation. (b) Correlation between the SFR per unit area obtained
   through H$\alpha$ and our synthesis (for $\tau = 24.5$ Myr). Units
   are $M_\odot\,yr^{-1}\,kpc^{-2}$ for both axis.  The dotted line
   marks the identity line.  (c) Correlation between the SFRs derived
   from equations (\ref{eq:SFR_synthesis}) and (\ref{eq:SFR_LHa}).
   Dashed lines indicate the $y(x)$ and $x(y)$ linear regressions,
   while the solid line shows the bisector fit.}
\label{fig:SFR_Ha_X_Synthesis}
\end{figure*}
%---------------------------- Figure ----------------------------

To compare the SFRs given by equations (\ref{eq:SFR_LHa}) and
(\ref{eq:SFR_synthesis}) we must first choose a specific value for
$\tau$. We do this by correlating the SFR {\em per unit area} obtained
with these two estimators, and seeking the value of $\tau$ which
yields the best correlation. Surface densities were used to remove the
$d^2$ factors common to both SFRs, thus avoiding distance-induced
correlations. Data for the SF$^{hq}$ sample was used in this
calibration. Also, since $L_{H\alpha}$ refers to the emission from
within the $3^{\prime\prime}$ SDSS fibers, $M^c_\star$ was not
extrapolated to the whole galaxy in this comparison.

Fig.\ \ref{fig:SFR_Ha_X_Synthesis} shows the results of this
exercise. Panel a shows the run of the Spearman coefficient ($R_S$)
for different values of $\tau$, with the best value indicated by an
arrow. Given the discreteness of our base, any value in the range of
the $t_\star = 25$ Myr bin yield identically strong correlations
(i.e., same $R_S$). We chose $\tau = 24.5$ Myr because this value
yields zero offset between these two SFRs. This is not a critical
choice, as values in the whole 10 to 100 Myr range yield correlations
of similar strength (Fig.\ \ref{fig:SFR_Ha_X_Synthesis}a). The
corresponding correlations between the synthesis and H$\alpha$-based
SFRs are shown in panels b and c in terms of SFR surface densities and
absolute SFRs, respectively. Robust fits to these relations yield
slopes very close to unity (1.09 in Fig.\
\ref{fig:SFR_Ha_X_Synthesis}b and 0.94 in Fig.\
\ref{fig:SFR_Ha_X_Synthesis}c).

The rms difference between these two SFR estimators is 0.3 dex,
corresponding to a factor of 2. We consider this an excellent
agreement, given that these estimators are based on entirely different
premises and independent data, and taking into account the
uncertainties inherent to both estimators. It is also reassuring that
$\tau$ turns out to be comparable to $t_{ion} \sim 10$ Myr, which is
(by construction) the smallest time-scale for SFR$_{H\alpha}$ to be
meaningful.  That the scatter between SFR$_{H\alpha}$ and ${\rm
SFR}_\star$ is typically just a factor of two can be attributed to the
fact that we are dealing with integrated galaxy data, thus averaging
over SF regions of different ages and emulating a globally constant
SFR, which works in the direction of compatibilizing the hypotheses
underlying equations (\ref{eq:SFR_LHa}) and (\ref{eq:SFR_synthesis}).

With these results, we define the ratio of ``current'' to mean past
SFR as

\begin{equation}
b = \frac{\overline{{\rm SFR}_\star}(\tau = 24.5 {\rm Myr})}
     {\overline{{\rm SFR}_\star}(\tau = \tau_G)}
\end{equation}

\ni where $\tau_G$ is the age of the oldest stars in a galaxy. In
practice, since the overwhelming majority of galaxies contain
components as old as our base allows, the denominator is simply
$M_\star^c$ divided by the age of the Universe, such that $b$ is
ultimately a measure of the current specific SFR. This definition
was used in Section \ref{sec:Correlations}, where its was shown that
$b$ decreases by an order of magnitude in the median from the top to
the bottom of the SF-wing.

%***************************************************************%
%                                                               %
%                          SFH: Results                         %
%                                                               %
%***************************************************************%

\section{The Star Formation Histories of SF Galaxies}
\label{sec:SFH_results}

Spectral synthesis methods such as the one employed in this work have
historically been seen with a good deal of skepticism, best epitomized
by \citet{Searle_1986}, who, when talking about the spectral synthesis
of integrated stellar populations, said that ``too much has been
claimed, and too few have been persuaded''. Persuading the reader that
one can nowadays recover a decent sketch of the time-dependent SFR in
a galaxy from its spectrum requires convincing results.  In this
section we apply the new tools to describe SFHs presented above
(equations \ref{eq:SFR} to \ref{eq:MAH}) to SF-galaxies in the
SDSS. As shown below, the temporal-dimension leads to a more detailed
view of SF-galaxies than that obtained with mean ages or current SFR
estimates.

\subsection{Distributions of Star Formation Histories}

%---------------------------- Figure ----------------------------
\begin{figure}
   \includegraphics[bb= 23 410 420 700, width=0.5\textwidth]
		  {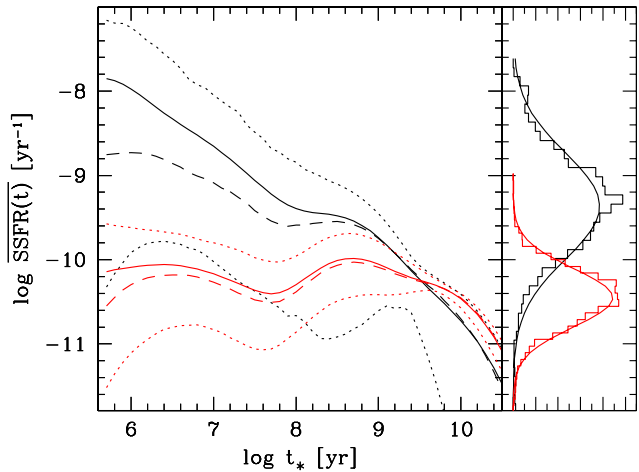}
\caption{{\em Left:} Distributions of star formation histories for
   $Z_{neb}$-bins A (black) and F (red), as defined in Fig
   \ref{fig:BPT}.  For each bin, we show the mean SSFR (solid
   line), the median (dashed) and the 5 and 95 percentiles of the
   distributions (dotted). {\em Right:} Normalized distributions of
   log SSFR for $t_\star = 25$ Myr. Gaussians are superimposed to
   illustrate that the distributions are log normal.}
\label{fig:SFH_distrib}
\end{figure}
%---------------------------- Figure ----------------------------

Our general strategy to explore the statistics of the sample is to
group galaxies according to certain similarity criteria and derive
mean SFHs for each group.  Since all the results presented from
Section \ref{sec:SFH_Zneb_bins} onwards are based on mean SFHs, it is
fit to first ask how representative such means are of the whole
distribution of SFHs.

This is done in Fig \ref{fig:SFH_distrib}, where we show the full
$t_\star$-by-$t_\star$ distribution of SSFR$(t_\star)$, computed with
equation (\ref{eq:SSFR}), for two of the six bins in $Z_{neb}$ defined
in Fig \ref{fig:BPT}: bins A and F, plotted in black and red, and
centered at $Z_{neb} = 0.31$ and 1.22, respectively. Solid lines
indicate the mean SSFR, dashed lines show the median and dotted lines
the corresponding 5 and 95 percentiles of the distributions. The first
thing one notices in this plot is that the distributions are very
wide.  For instance, for most of the $t_\star < 1$ Gyr range, their 5
to 95 percentile ranges span over 2 orders of magnitude in SSFR. As
discussed further below, this is in part due to the choice of grouping
galaxies by $Z_{neb}$. Grouping by properties more directly related to
the SFHs should lead to narrower distributions. However, one must
realize that since galaxy evolution depends on many factors, grouping
objects according to any single property will {\em never} produce
truly narrow SFH distributions. Secondly, the distribution of SSFR
values at any $t_\star$ is asymmetric, as can be seen by the
fact that the mean and median curves differ. In fact, as illustrated
by the right panel in Fig \ref{fig:SFH_distrib}, these distributions
are approximately {\em log-normal}, indicating that SFHs result from
the multiplication of several independent factors, as qualitatively
expected on physical grounds.

Despite their significant breadth and overlap, it is clear that the
SSFR-distributions for $Z_{neb}$ bins A and F in Fig
\ref{fig:SFH_distrib} are very different, particularly at low
$t_\star$. This is confirmed by KS tests, which show that these two
distributions are undoubtedly different.  In fact, for {\em any} pair
of $Z_{neb}$ bins the distributions differ with $> 99\%$ confidence
for {\em any} $t_\star$,

In what follows we will present only mean SFHs, obtained grouping
galaxies according to a subset of the available physical parameters.
Whilst their is clearly more to be learned from the SFH distributions
discussed above, this is a useful first approach to explore the
intricate relations between galaxy properties and their SFHs.

\subsection{Trends along the SF-wing}
\label{sec:SFH_Zneb_bins}

%---------------------------- Figure ----------------------------
\begin{figure}
   \includegraphics[bb= 40 270 340 695, width=0.5\textwidth]
		  {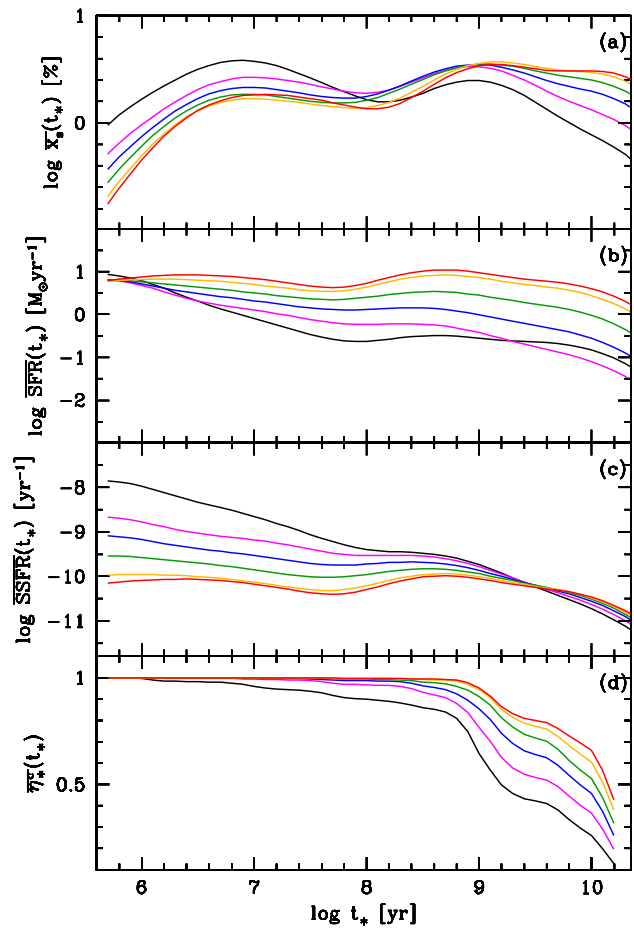}
\caption{Mean star formation histories for the 6 different
   $Z_{neb}$-bins defined in Fig \ref{fig:BPT}, in four different
   representations: (a) smoothed population vector,
   $\overline{x_s}(t_\star)$, (b) $\overline{\rm SFR}(t_\star)$, (c)
   $\overline{\rm SSFR}(t_\star)$, and (d)
   $\overline{\eta^c_\star}(t_\star)$.}
\label{fig:mean_SFHs}
\end{figure}
%---------------------------- Figure ----------------------------

We start our statistical study of galaxy SFHs grouping galaxies in the
six $Z_{neb}$ bins defined in Section \ref{sec:BPTdiagram}. As shown
in Fig \ref{fig:BPT}, $Z_{neb}$ traces the location of a galaxy along
the SF-wing in the BPT diagram.

Fig \ref{fig:mean_SFHs} shows the derived SFHs for the A--F bins in
four different representations, from top to bottom: $x_s$, SFR, SSFR
and $\eta^c_\star$ as a function of stellar age $t_\star$. Each line
represents a $t_\star$-by-$t_\star$ average over all galaxies in the
bin. The plots show that young populations are present in a proportion
which increases systematically as $Z_{neb}$ decreases. This is
evident, for instance, in the $x_s(t\star)$ panel, which shows how
$Z_{neb}$-related age distributions combine to produce the correlation
between $\langle \log t_\star \rangle_L$ and $Z_{neb}$ depicted in
Fig.~\ref{fig:ZnebCorrelations}f.

The SFR$(t\star)$ curves (panel b) show that SF-galaxies of different
$Z_{neb}$ differ more in their past SFR, low $Z_{neb}$ having SFRs
about 100 times lower than those of high $Z_{neb}$ a few Gyr ago. In
the more recent past, all curves converge to SFRs of a few
$M_\odot/yr$. At first sight, this convergence seems to be at odds
with the fact that galaxies with low and high $Z_{neb}$ differ by
about one order of magnitude in the median $L_{H\alpha}$ (Table
\ref{tab:ZnebBinsStats}), and thus should differ by a similar factor
in the recent SFR. In fact, there is no contradiction, since what
needs to be considered when comparing the synthesis-based SFR with
that derived from H$\alpha$ is the mean SFR over scales of at least 10
Myr, and these are clearly smaller for low $Z_{neb}$ galaxies than for
those of higher $Z_{neb}$. As shown in
Fig~\ref{fig:SFR_Ha_X_Synthesis}, H$\alpha$ and synthesis based SFRs
agree very well.  Ultimately, the apparent coincidence of mean SFR
curves of different $Z_{neb}$ is due to the fact that the relation
between recent SFR and $Z_{neb}$ is a relatively weak and scattered
one (Fig.~\ref{fig:ZnebCorrelations}h), such that along the whole
SF-wing one may find galaxies that transform a similar amount of gas
into stars per year.

The clearest separation between SFHs of galaxies of different
$Z_{neb}$ is provided by the SSFR$(t\star)$ curves. At ages $\ga$ a
few Gyr all SSFR curves merge. This behavior is a consequence of the
fact that most of the stellar mass is assembled early on in a galaxy's
history, irrespective of $Z_{neb}$ or other current properties. With
our $\Delta \log t_\star = 1$ dex smoothing, this initial phase, over
which $\int {\rm SFR} dt_\star \sim M^c_\star$, becomes a single
resolution element in the SFR curves. Division by $M_\star^c$ to
produce a specific SFR (equation \ref{eq:SSFR}) then makes all curves
coincide.  At later times (smaller $t_\star$), however, the curves
diverge markedly, with the lowest and highest $Z_{neb}$ groups
differing in SSFRs by $\sim 2$ orders of magnitude nowadays. This
confirms that the relation between recent and past star-formation is a
key-factor in distributing galaxies along the SF-wing in the BPT
diagram (Fig.~\ref{fig:ZnebCorrelations}g).

Yet another way to visualize the SFH is through the mass-assembly
function defined in equation (\ref{eq:MAH}).  Though
$\eta^c_\star(t_\star)$ is computed with the raw (unsmoothed)
population vector, for presentation purposes we apply a FWHM $= 0.2$
dex gaussian in $\log t_\star$, just enough to smooth discontinuities
associated with the discrete set of $t_{\star,j}$'s in our base.  Fig
\ref{fig:mean_SFHs}d shows the results.  This is essentially a
cumulative representation of the same results reported in
Fig\ref{fig:mean_SFHs}c, namely, that low $Z_{neb}$ galaxies are
slower in assembling their stars. This plot is however better than
the previous ones in showing that despite these differences, all
galaxies have built up most of their stellar mass by $t_\star = 1$ Gyr.

%***************************************************************%
%                                                               %
%                           Mass Bins                           %
%                                                               %
%***************************************************************%

These encouraging results indicate that synthesis methods have evolved
to a point where one can use them in conjunction with the fabulous
data sets currently available to sketch a fairly detailed
semi-empirical scenario for galaxy evolution. In the next section we
walk a few more steps in this direction by inspecting how
astrophysically plausible drivers of galaxy evolution relate to the
SFHs recovered from the data.

\subsection{Star Formation Histories and Chemical Evolution in Mass and
Surface-Density bins}
\label{sec:MassBins}

The value of grouping galaxies by $Z_{neb}$ is that it maps SFHs to a
widely employed diagnostic tool: the BPT diagram.  Yet, present day
nebular abundance is not a cause, but a consequence of galaxy
evolution.  In this section we leave aside our focus on the BPT
diagram and group galaxies according to properties more directly
associated to physical drivers of galaxy evolution. Two natural
candidates are the mass ($M_\star$) and surface mass-density
($\Sigma_\star$).  Like $Z_{neb}$, both $M_\star$ and $\Sigma_\star$
can be considered the end product of a SFH, yet they are clearly more
direct tracers of depth of the potential well and degree of gas
compression, two key parameters affecting physical mechanisms which
regulate galaxy evolution (\citealp{Schmidt_1959,
Tinsley_1980,Kennicutt_1998}).

\begin{figure}
   \includegraphics[bb= 40 270 340 695, width=0.5\textwidth]
		  {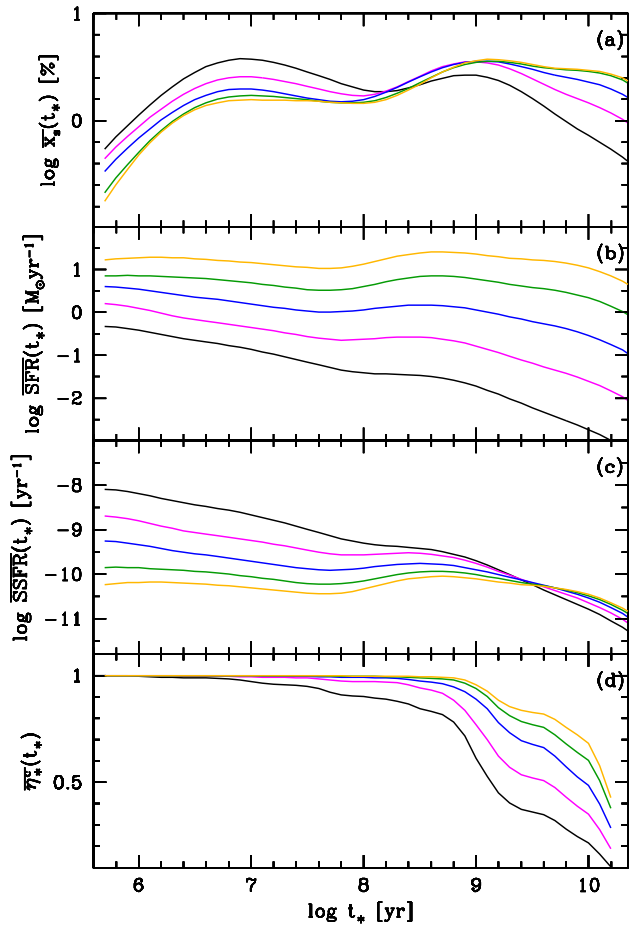}
\caption{As Fig \ref{fig:mean_SFHs}, but binning SF galaxies by their
   stellar mass, using five 1 dex wide bins, centered at (from bottom
   to top in panel b) $\log M_\star/M_\odot = 7.5$, 8.5, 9.5, 10.5 and
   11.5, which contain 542, 4057, 26700, 47153, 3808 galaxies,
   respectively.}
\label{fig:mean_SFHs_MassBins}
\end{figure}

Fig \ref{fig:mean_SFHs_MassBins} shows our different representations
of the SFH of SF galaxies for five 1 dex-wide mass bins centered at
$\log M_\star/M_\odot = 7.5 \cdots 11.5$.  Given that $M_\star$ and
$Z_{neb}$ are related, the overall evolutionary picture emerging from
this plot is similar to the one obtained binning galaxies in
$Z_{neb}$, i.e., massive galaxies assemble their stars faster than
low-mass galaxies. On the whole, Fig \ref{fig:mean_SFHs_MassBins}
provides a compelling visualization of galaxy downsizing.

The most noticeable difference with respect to $Z_{neb}$-binned
results is on the absolute SFR curves (compare Figs
\ref{fig:mean_SFHs}b and \ref{fig:mean_SFHs_MassBins}b). This
difference is rooted in the fact that galaxies of similar $Z_{neb}$
span a much wider range of SFRs than galaxies of similar $M_\star$.
This can be illustrated focusing on recent times, and inspecting the
$L_{H\alpha}$--$Z_{neb}$ relation (Fig.\ \ref{fig:ZnebCorrelations}h),
with the understanding that $L_{H\alpha}$ can be read as the current
SFR (Fig.\ \ref{fig:SFR_Ha_X_Synthesis}). Despite the statistically
strong correlation ($R_S = 0.49$), the typical 5--95 percentile range
in $L_{H\alpha}$ for a given $Z_{neb}$ is $\sim 2.1$ dex, comparable
to the full dynamic range spanned by the data (2.4 dex over the same
percentile range). In other words, the relation has a large scatter,
and hence $Z_{neb}$-binning mixes objects with widely different SFRs,
explaining why the $\overline{\rm SFR}(t_\star)$ curves in Fig
\ref{fig:mean_SFHs}b tend to merge at low $t_\star$.  The
$L_{H\alpha}$-$M_\star$ relation (not shown), on the other hand, is
stronger ($R_S = 0.68$), partly due to the $d^2$ factors in common to
absolute SFRs and $M_\star$. Grouping by $M_\star$ then selects
galaxies in narrower SFR ranges, producing the well separated curves
seen in Fig \ref{fig:mean_SFHs_MassBins}b.

%---------------------------- Figure ----------------------------
%---------------------------- Figure ----------------------------

\begin{figure}
   \includegraphics[bb= 40 270 340 695, width=0.5\textwidth]
		  {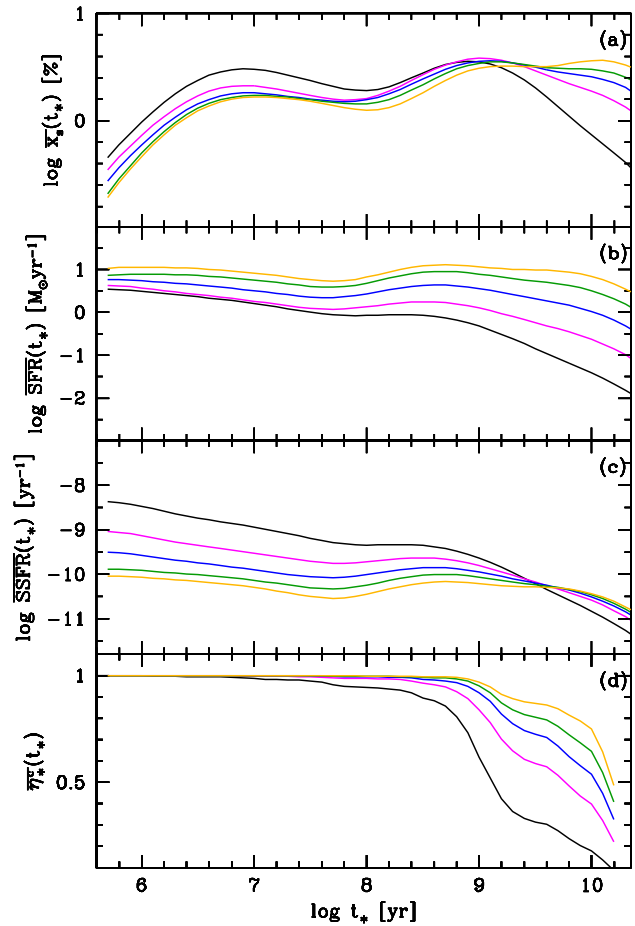}
\caption{As Fig \ref{fig:mean_SFHs}, but binning SF galaxies by their
stellar surface densities, with five 0.5 dex wide bins centered at
$\log \Sigma_\star = 7.25$, 7.75, 8.25, 8.75 and 9.25
$M_\odot\,$kpc$^{-2}$, containing 1477, 12554, 37177, 27742, 3046
galaxies respectively.}
\label{fig:mean_SFHs_SurfDenBins}
\end{figure}

Results grouping galaxies according to $\Sigma_\star$ are presented in
Fig \ref{fig:mean_SFHs_SurfDenBins}. Since $\Sigma_\star$ and
$M_\star$ correlate very strongly ($R_S = 0.73$ in our sample), the
results are similar to those obtained grouping galaxies by their
stellar mass. \citet{Kauffmann_etal_2003b}, based on an analysis of
two SFH-sensitive spectroscopic indices (namely $D_n(4000)$ and
H$\delta_A$), propose that $\Sigma_\star$ is more directly connected
to SFHs than $M_\star$. This is not obviously so comparing Figs
\ref{fig:mean_SFHs_MassBins} and \ref{fig:mean_SFHs_SurfDenBins}. A
more detailed, multivariate analysis is needed to evaluate which is
the primary driver of SFHs.

%---------------------------- Figure ----------------------------
%---------------------------- Figure ----------------------------

%***************************************************************%
%                                                               %
%                        Sample Selection                       %
%                                                               %
%***************************************************************%

%\section{Alternative sample selections}
%\section{Analysis of possible bias}

\section{Selection effects and modelling caveats}
\label{sec:samples}

This section deals with the effects of sample selection, synthesis
ingredients and model assumptions on our results.

\subsection{Selection effects}

We now study to which extent the mean SFHs derived in the last section
are affected by the way we have defined SF galaxies.  We address this
issue recomputing mean SFHs for samples constructed with alternative
selection criteria, and comparing to the results obtained with our
default sample.

%---------------------------- Figure ----------------------------
\begin{figure*}
   \includegraphics[bb=40 460 572 700,width=\textwidth]
		  {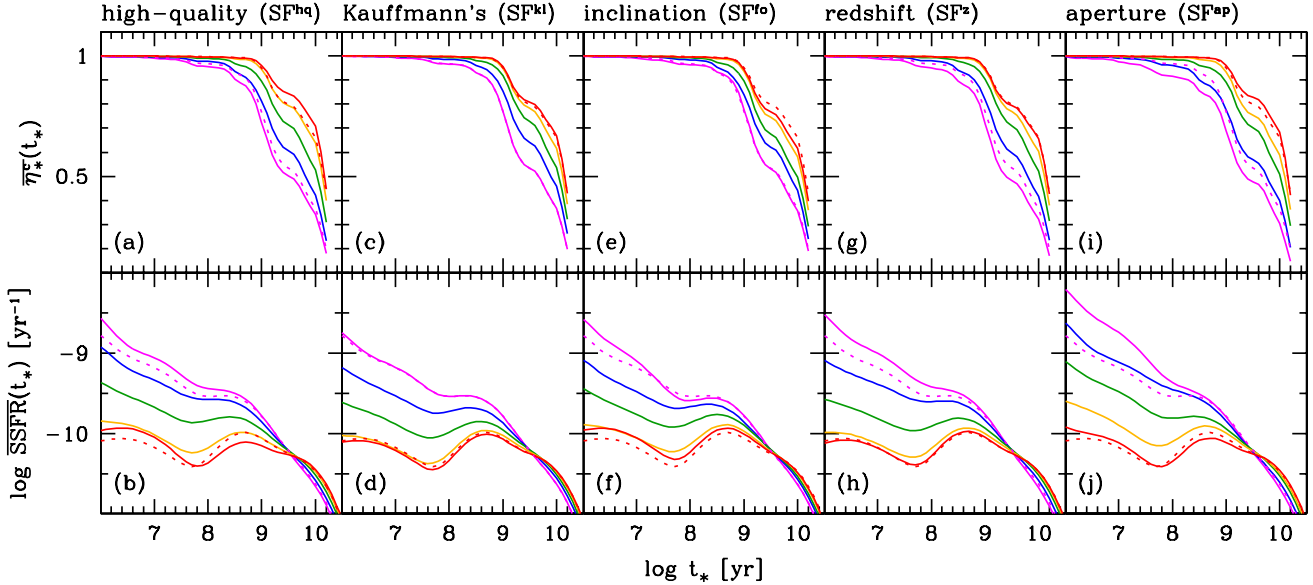}
   \caption{Average mass assembly histories ($\eta^c_\star$, top
     panels) and specific star formation rate (${\rm SSFR}$, bottom)
     histories for $Z_{neb}$-bins B to F, color-coded as in Fig
     \ref{fig:BPT}. Solid lines show the curves for different sample
     definitions.  For comparison, the dotted lines in all panels show
     the curves for bins B (lower curves; magenta) and F (upper
     curves; red) of the full SF sample.}
\label{fig:alt_samples}
\end{figure*}
%---------------------------- Figure ----------------------------

We first ask how our emission line and continuum $S/N$ cuts influence
our results.  Figs \ref{fig:alt_samples}a and b show the average mass
assembly and SSFR functions for the high-quality SF$^{hq}$ sub-sample
defined in Section \ref{sec:sample_definition}. The results for this
better-data sub-sample are very similar to those obtained with the
full SF sample. The SSFR curves in recent times are skewed to slightly
higher rates, which reflects the fact that objects in the SF$^{hq}$
sample are slightly younger than those in the full SF sample, as shown
in Fig \ref{fig:obs_prop}.

Our BPT-based selection of SF galaxies used the dividing line proposed
in SEAGal III, which is more restrictive than the empirical line
proposed by \citet{Kauffmann_etal_2003c}. We define the SF$^{kl}$
sample as the 111026 galaxies classified as SF according to the
\citet{Kauffmann_etal_2003c} line.  Fig.~\ref{fig:alt_samples}c and d
show that the SFHs for $Z_{neb}$ bins in this sample are nearly
indistinguishable from those obtained with the SEAGal classification
criterion.

Another concern is the inclination effect.  The spectra of edge-on
objects are biased by the metal poorer outer parts of the galaxies,
leading to an underestimation of $Z_{neb}$. This may lead us to place
a galaxy in a lower $Z_{neb}$ bin than it would if it were seen face
on, possibly affecting the mean SFH in that bin.  To investigate this
effect we have defined a sub-sample of nearly face-on galaxies,
SF$^{fo}$ (6842 objects), selecting by the inclination parameter, $b/a
\ge 0.9$. Figs~\ref{fig:alt_samples}e and f show that the SFHs derived
with this sample are practically the same as for the full sample.

Aperture effects are a common source of concern in studies of SDSS
spectra (e.g., \citealp{Gomez_etal_2003}; SEAGal I).  To investigate
how such effects impact upon our SFHs we defined two samples: the
SF$^{z}$ sample, which comprises 58153 SF galaxies with $z \ge 0.05$
(as opposed to 0.002 for the full SF sample), and the SF$^{ap}$
sample, comprising only the 1096 objects with more than half of their
$z$ band luminosity inside the fiber.  Both criteria exclude
proportionately more the population of low $Z_{neb}$, low $M_\star$
galaxies (distant galaxies of this category are not present in the
SDSS, due to limiting magnitude).  Accordingly, the change in SFHs is
only noticeable for the lowest $Z_{neb}$ bins, as shown in
Figs~\ref{fig:alt_samples}g--j. In particular,
Fig~\ref{fig:alt_samples}j shows that the mean SFH for bin C in the
SF$^{ap}$ sample matches that of bin B in the full sample. Of all
selection-induced changes discussed here, this is the largest one;
yet, all it does is to shift the SFHs from one group of galaxies to
the adjoining one.

To summarize, selection criteria may influence the derived mean SFHs
in quantitative terms, but do {\em not} modify the relative pattern of
mean SFHs of galaxies in different $Z_{neb}$ bins. The same applies to
grouping galaxies according to properties other than $Z_{neb}$. The
general trends in the SFH as a function of global galaxy properties
obtained in this work are therefore robust against variations in the
sample selection criteria.

\subsection{Experiments with different models}

One should also ask to which extent our results are robust against
changes in the base of evolutionary synthesis models. While answering
this question requires an in-depth study far beyond the scope of this
paper, we believe this has a much larger impact on SFHs than selection
effects.

Panter \etal (2007) reported results of MOPED experiments using
spectral models from different sources (\citealp{Jimenez_etal_2004,
Fioc_Rocca-Volmerange_1997, Maraston_2005, Bruzual_Charlot_1993};
and BC03), all sampled at $\Delta \lambda = 20$ \AA. For the 767
galaxies in their randomly selected test sample, the resulting mean
star-formation fractions (analogous to our $\vec{\mu}$ vector) differ
by factors of a few for the youngest and oldest ages, and close to a
full order of magnitude for $t_\star \sim 0.1$--1 Gyr. Recovering SFHs
in this intermediate age range is particularly hard, as discussed by
\citet{Mathis_Charlot_Brinchmann_2006}. Indeed, the experiments
reported by Panter \etal often find a suspiciously large 1 Gyr
component.

The behaviour of our \starlight\ fits is also anomalous in this age
range. This is clearly seen in the mean SFHs shown in Figs
\ref{fig:mean_SFHs}--\ref{fig:mean_SFHs_SurfDenBins}, particularly in
the $\vec{x}_s$ representation, which shows a hump at $\sim 1$
Gyr. Interestingly, \starlight\ experiments with a base of
evolutionary synthesis models using the MILES library of
\citet[instead of the STELIB library used in the base adopted for
this and previous SEAGal studies]{Sanchez-Blazquez_etal_2006} do not
produce this hump at $\sim 1$ Gyr. In fact, the whole mean SFHs
derived with this new set of models differs systematically from those
shown in Figs
\ref{fig:mean_SFHs}--\ref{fig:mean_SFHs_SurfDenBins}. The mass
assembly function $\etaÄ_\star(t_\star)$, for instance,
rises more slowly and converges at somewhat smaller $t_\star$ than
those obtained in this work. There are also systematic differences in
stellar extinction, which comes out $\Delta A_V^\star \sim 0.4$ mag
larger with the MILES models. Extensive tests with these new bases,
including different prescriptions of stellar evolution as well as
different spectral libraries, are underway, but these first results
show that significant changes can be expected.

Reassuringly, however, these same experiments show that the pattern of
SFHs as a function of $Z_{neb}$, $M_\star$ and $\Sigma_\star$ reported
in this paper does {\em not} change with these new models. In any
case, these initial results provide an eloquent reminder of how
dependent semi-empirical SFH studies are on the ingredients used in
the fits.

\subsection{Experiments with differential extinction}
\label{sec:samples_YAV}

%---------------------------- Figure ----------------------------
\begin{figure}
  \includegraphics[bb= 40 180 340 600, width=0.5\textwidth]
		  {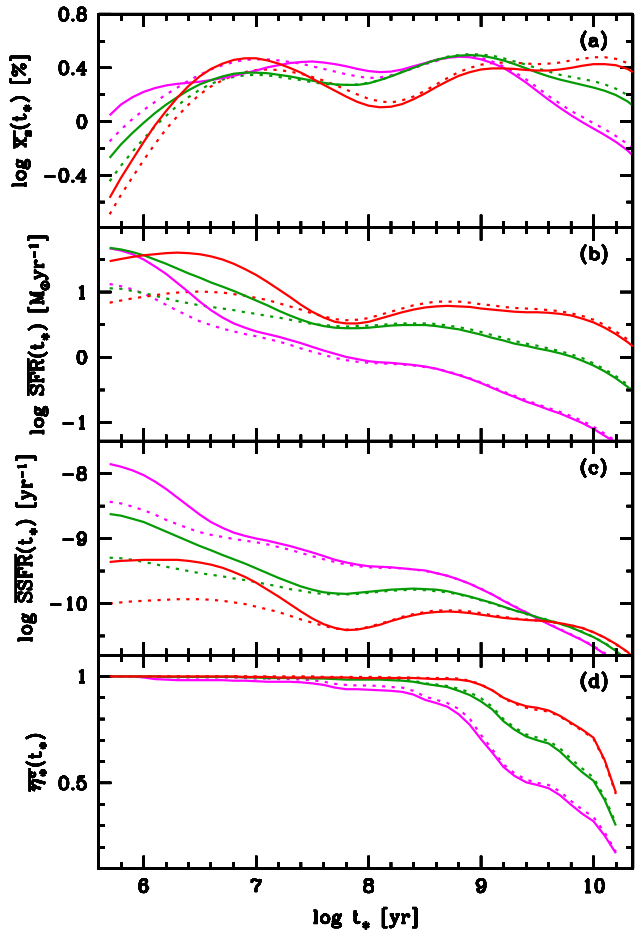}
\caption{Comparison between mean star formation histories for the
  SF$^{hq}$ sample modelled with only one stellar extinction (dotted
  lines) and with differential extinction (solid lines).  For clarity,
  only $Z_{neb}$-bins B, D and F are drawn, and color-coding is the
  same as in Fig \ref{fig:BPT}. Each panel shows a different
  representation of SFHs: (a) smoothed population vector,
  $\overline{x_s}(t_\star)$, (b) $\overline{\rm SFR}(t_\star)$, (c)
  $\overline{\rm SSFR}(t_\star)$, and (d)
  $\overline{\eta^c_\star}(t_\star)$.}
\label{fig:SFHs_17kYAV}
\end{figure}
%---------------------------- Figure ----------------------------

We saw in Section \ref{sec:av_neb} that the nebular and stellar
exctintions are strongly correlated, but with $A_V^{neb} \sim$ 
twice $A_V^{\star}$.  This indicates that the uniform stellar
extinction used in our fits is not adequate to model star-forming
regions, which should be subjected to a similar extinction than the
line-emiting gas, ie., $A_V^\star(t_\star \la 10^7 {\rm yr}) \sim
A_V^{neb}$.  It is therefore fit to ask whether and how such a
differential extinction affects our general results.

Given the difficulties in recovering reliable population dependent
$A_V^\star$'s from spectral synthesis in the optical range alone, we
address this question by postulating that populations younger than
$10^7$ yr are extincted by $0.34 + 2.28 A_V^{\star}$, i.e., the
empirical $A_V^{neb}(A_V^\star)$ relation found in Section
\ref{sec:av_neb}, with $A_V^{\star}$ now denoting the extinction to $>
10^7$ yr stars.\footnote{We thank the referee for suggesting this
approach.}  The 17142 galaxies in the SF$^{hq}$ sub-sample were
re-fitted with this more realistic modified recipe for extinction
effects.

Qualitatively, one expects that forcing a uniform $A_V^\star$ fit to a
galaxy where the young stars suffers more extinction than the others
should lead to an overestimation of the age of the young population.
This older and thus redder young population would compensate the
mismatch in $A_V^\star$.  Allowing for $A_V^\star(t < 10^7~{\rm yr})$
larger than the $A_V^\star$ of the $t > 10^7~{\rm yr}$ populations
makes it possible to fit the same spectrum with younger and dustier
populations. Hence, the recent SFRs should increase.

This expectation is fully confirmed by these new fits. Fig.
\ref{fig:SFHs_17kYAV} shows a comparison between three $Z_{neb}$ bins
(B, D and F) for the old and new fits of the SF$^{hq}$ sample.  One
sees that the new average SFR and SSFR curves are shifted by $\sim
0.3$ dex upwards in the $t_\star \le 10^7$ yr range with respect to
the ones obtained with a single extinction. The rearrangements in the
population vector tend to be in the sense of shifting some light from
old  populations to the $\le 10^7$ yr components.

Not surprisingly, the properties which change most are those directly
related to the strength of the young population, such as current SFR,
which increases by 0.3 dex on average, and the mean stellar age, which
decreases by $\Delta \langle \log t_\star \rangle_L \sim 0.1$ dex. The
changes in other global properties such as $A_V^\star$ and $M_\star$
are much smaller than this.

These experiments are obviously a simplification of the problem of
dust distribution in galaxies, yet they suggest that the choice of the
extinction modelling can have non-negligible effects on the derived
SFH curves.  On the whole, however, the qualitative pattern of SFHs as
a function of $Z_{neb}$ or other variables stays the same. As found in
the sample selection studies, quantitative changes are at best
equivalent to moving from one bin to the next, so that our general
conclusion does not change.

%***************************************************************%
%                                                               %
%                           Summary                             %
%                                                               %
%***************************************************************%

\section{Summary}
\label{sec:Summary}

In this paper we have studied physical properties of 82302 normal
star-forming galaxies from the SDSS DR5, by using results from our
stellar population synthesis code, \starlight, and our emission-line
measurement algorithm.

Before reviewing our main results, we highlight some aspects of this
study which have relatively little impact upon our general
conclusions, but represent significant refinements in our methodology.

\begin{enumerate}

\item We have detected a systematically overestimated continuum level
around H$\beta$, whose origin we tentatively attribute to deficiencies
in the STELIB calibration in this range. Gaussian fits to the H$\beta$
emission which disregard this offset tend to underestimate the line
flux by 4\% on average, and $\sim 7$\% in the case of weaker lines.
These are relatively small, yet systematic effects, which propagate to
estimates of nebular extinction, metallicities and galaxy
classification.

\item SF galaxies were selected according to the theoretical
criterion proposed in SEAGal III, which minimizes contamination
by AGN emission.

\item Nebular extinctions and metallicities were derived
self-consistently, allowing for the metallicity dependence of the
Balmer decrement. Five different reddening laws were explored, but
found to produce equally good spectral fits and relatively small
differences in derived physical properties.

\item We have confirmed the strong correlation between $A_V^{neb}$ and
$A_V^\star$ found in SEAGal I. We have also identified a strong
correlation between the strength of the ISM component of the NaD
absorption doublet and the amount of dust derived from the synthesis.

\item Different recipes for nebular metallicity estimates were tried.
Some of them proved not to be adequate for this study, either because
of the lack of spectral data (e.g., measures of [ArIII]$\lambda$7135
and [OIII]4363 emission lines were available for few objects), or
because such calibrations were only valid in the low-$Z_{neb}$ regime,
thus encompassing a very small fraction of objects from our sample.
Therefore, throughout our analysis we use the O$_3$N$_2$ index and the
calibration by \citet{Stasinska_2006} to measure the nebular
metallicity. Although this is not a reliable calibrator at the lowest
metallicities, it is good enough for our analysis in
$Z_{neb}$-bins. Furthermore, it has the nice virtue of being directly
related with the position of the objects in the BPT diagram.

\end{enumerate}

We now summarize results related to the main goal of this paper,
namely, to investigate the SFH of galaxies along the SF wing in the
BPT diagram.  In practice, this means studying how SFHs change as a
function of nebular metallicity, even though $Z_{neb}$ is more a
product than a cause of galaxy evolution.

\begin{enumerate}

\item We started our study with a traditional analysis, correlating
$Z_{neb}$ with several physical and observed properties. This analysis
confirms results obtained directly or indirectly in the past by other
works, such as relations between the nebular metallicity and galaxy
luminosity, mass, dust content, mean stellar metallicity and mean
stellar age.

\item Formalism towards a time-dependent analysis was then
presented. Simple ways to compress the output of our stellar
population synthesis code were proposed. These are based either on a
posteriori smoothing of the age distribution, which allows the
derivation of time dependent star formation rates, or a cumulative
mass assembly history.

\item As a first application of this time dependent description of
SFHs we computed the current SFR obtained from our spectral fits.  The
resulting values of SFR$_\star$ agree very well with more traditional
estimates based on the luminosity of H$\alpha$.  The scatter between
SFR$_\star$ and SFR$_{H\alpha}$ is just a factor of 2, despite the
differences in the underlying assumptions and sensitivity to the IMF.
This result strengthens confidence in our method, and, more
importantly, opens the possibility of measuring current SFRs in
galaxies hosting AGN, where H$\alpha$ method does not apply.

\item Fully time dependent SFHs were then derived grouping galaxies
into six $Z_{neb}$ bins spanning the entire SF wing of the BPT
diagram.  Mean SFHs for each of these bins were presented in four
different representations: (a) the smoothed population vector,
$\overline{x_s}(t_\star)$, (b) the star formation rates
$\overline{\rm SFR}(t_\star)$, (c) specific star formation rates
$\overline{\rm SSFR}(t_\star)$, and (d) mass-assembly histories ,
$\overline{\eta_\star^c}(t_\star)$.

\item We found that SFHs vary systematically along the SF
sequence.  Though all galaxies assembled the bulk of their stellar
mass over 1 Gyr ago, low $Z_{neb}$ systems evolve at a slower pace.
Galaxies at the tip of the SF wing have current specific SFRs about 2
orders of magnitude larger than the metal rich galaxies at the the
bottom of the BPT diagram.

\item At any given time, the distribution of SSFRs for galaxies within
a $Z_{neb}$-bin is quite broad and approximately log-normal.

\item We performed the same SFH study grouping galaxies by their
stellar mass and surface mass density. Given the existence of
$Z_{neb}$--$M_\star$--$\Sigma_\star$ relations, the overall picture is
obtained grouping by $Z_{neb}$. Thus, low $M_\star$ (low
$\Sigma_\star$) systems are the ones which evolve slower, with current
SSFRs much larger than more massive (dense) galaxies.

\item Finally, we have analysed a number of selection and modelling
effects that might bias our results, and show that while they may
affect the derived SFHs quantitatively, the organization of SFHs as a
function of $Z_{neb}$, $M_\star$, $\Sigma_\star$ remains the same.
Experiments with new evolutionary synthesis models and differential
extinction fits were reported and
found to lead to substantially different SFHs, yet preserving this
same overall pattern.

\end{enumerate}

%***************************************************************%
%                                                               %
%                       Acknowledgments                         %
%                                                               %
%***************************************************************%

\section*{ACKNOWLEDGMENTS}

We are greatly in debt with several colleagues and institutions around
the globe who have contributed to this project by allowing access to
their computers.  The \starlight\ project is supported by the
Brazilian agencies CNPq, CAPES, FAPESP, by the France-Brazil
CAPES/Cofecub program and by Observatoire de Paris.

Funding for the SDSS and SDSS-II has been provided by the Alfred
P. Sloan Foundation, the Participating Institutions, the National
Science Foundation, the U.S. Department of Energy, the National
Aeronautics and Space Administration, the Japanese Monbukagakusho, the
Max Planck Society, and the Higher Education Funding Council for
England. The SDSS Web Site is http://www.sdss.org/. The SDSS is
managed by the Astrophysical Research Consortium for the Participating
Institutions. The Participating Institutions are the American Museum
of Natural History, Astrophysical Institute Potsdam, University of
Basel, University of Cambridge, Case Western Reserve University,
University of Chicago, Drexel University, Fermilab, the Institute for
Advanced Study, the Japan Participation Group, Johns Hopkins
University, the Joint Institute for Nuclear Astrophysics, the Kavli
Institute for Particle Astrophysics and Cosmology, the Korean
Scientist Group, the Chinese Academy of Sciences (LAMOST), Los Alamos
National Laboratory, the Max-Planck-Institute for Astronomy (MPIA),
the Max-Planck-Institute for Astrophysics (MPA), New Mexico State
University, Ohio State University, University of Pittsburgh,
University of Portsmouth, Princeton University, the United States
Naval Observatory, and the University of Washington.

%***************************************************************%
%                                                               %
%                          References                           %
%                                                               %
%***************************************************************%

\end{document}